%% file: main.tex
\documentclass[10pt,sigconf,letterpaper,nonacm,balance=false]{acmart}
\input{sections/packages}
\input{sections/numbers}
\input{sections/numbers_imc_text}   
\IfFileExists{sections/numbers_imc_engine.tex}{\input{sections/numbers_imc_engine}}{}
\IfFileExists{sections/numbers_imc_supply.tex}{\input{sections/numbers_imc_supply}}{}
\input{sections/numbers_imc_fixes}    



\acmConference[IMC '27]{Proceedings of the 2027 ACM Internet Measurement Conference}{2027}{}
\acmBooktitle{Proceedings of the 2027 ACM Internet Measurement Conference (IMC '27)}
\acmYear{2027}
\copyrightyear{2027}
\setcopyright{none}
\acmDOI{}
\acmISBN{}

\begin{document}

\title{\toolstyle{GEOHunter}: Mapping a Generative Engine Optimization Network in the Wild}


\author{
  Sean Tang$^1$, Peiran Wang$^1$, Ying Li$^1$, Esha Shivakumar$^1$, Tao Meng$^2$, Yuan Tian$^1$
}
\affiliation{%
  \institution{
    $^1$University of California, Los Angeles (UCLA) \quad
    $^2$Georgia Institute of Technology
  }
  \country{}
}


\renewcommand{\shortauthors}{Anonymous et al.}

\input{sections/abstract}



\maketitle

\input{sections/1-introduction}
\input{sections/2-background}
\input{sections/3-mapping}
\input{sections/4-rq1-organization}
\input{sections/6-rq3-assistants}
\input{sections/8-related}      
\input{sections/7-discussion}
\input{sections/9-conclusion}     


\newpage\clearpage   
\bibliographystyle{ACM-Reference-Format}
\bibliography{references}

\appendix
\input{sections/appendix-imc}
\input{sections/appendix}

\end{document}

%% file: sections/packages.tex
\usepackage{multirow}
\usepackage{subcaption}

\usepackage{iftex}
\ifXeTeX\else
  \PackageError{packages}{This draft must be compiled with XeLaTeX}{On Overleaf: Menu > Settings > Compiler > XeLaTeX
    (latexmkrc should already select it). Locally: ./build.sh}
\fi
\IfFontExistsTF{Noto Serif CJK SC}
  {\newfontfamily\cjkfont{Noto Serif CJK SC}}
  {\newfontfamily\cjkfont{FandolSong}[Extension=.otf, UprightFont=*-Regular, BoldFont=*-Bold]}
\XeTeXlinebreaklocale "zh"
\newXeTeXintercharclass\cjkclass
\makeatletter
\def\cjk@range#1#2{\count@=#1\relax\loop\ifnum\count@<#2\relax
  \XeTeXcharclass\count@=\cjkclass\advance\count@ by 1\repeat}
\cjk@range{"3000}{"3040}  
\cjk@range{"4E00}{"A000}  
\cjk@range{"FF00}{"FFF0}  
\makeatother
\XeTeXinterchartoks 0 \cjkclass = {\begingroup\cjkfont}
\XeTeXinterchartoks 4095 \cjkclass = {\begingroup\cjkfont}
\XeTeXinterchartoks \cjkclass 0 = {\endgroup}
\XeTeXinterchartoks \cjkclass 4095 = {\endgroup}

\newif\ifdraft\draftfalse
\ifdraft
  \newcommand{\todo}[1]{{\color{red}[\textbf{TODO:} #1]}}
  
  \newcommand{\note}[1]{{\color{blue}[\textit{#1}]}}
  \definecolor{answerblue}{RGB}{0,70,200}
  
\else

  \newcommand{\todo}[1]{}

  \newcommand{\note}[1]{}
\fi

\usepackage{xspace}
\DeclareRobustCommand{\toolstyle}[1]{\textsc{#1}}
\DeclareRobustCommand{\geohunter}{\toolstyle{GEOHunter}\xspace}
\DeclareRobustCommand{\geoprobe}{\toolstyle{GEOProbe}\xspace}
\usepackage[framemethod=tikz]{mdframed}
\newcommand{\finding}[2]{\begin{mdframed}[nobreak=true,linewidth=0.4pt,linecolor=black,innertopmargin=3pt,innerbottommargin=3pt,innerleftmargin=4pt,innerrightmargin=4pt,skipabove=\smallskipamount,skipbelow=\smallskipamount,splittopskip=6pt]\noindent\textbf{Finding #1.} #2\end{mdframed}}

\makeatletter
\def\@parfont{\bfseries}
\makeatother

\definecolor{curblue}{RGB}{0,70,200}   

\usepackage{enumitem}
\usepackage{tikz}
\usetikzlibrary{arrows.meta,calc,positioning,fit,backgrounds}

\usepackage{colortbl}
\newcommand{\iconbox}[1]{\raisebox{-0.17\height}{#1}}
\newcommand{\iconBili}{\iconbox{\includegraphics[height=0.95em]{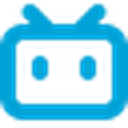}}}
\newcommand{\iconBjh}{\iconbox{\includegraphics[height=0.95em]{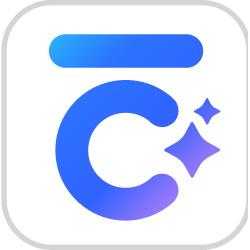}}}
\newcommand{\iconKuai}{\iconbox{\includegraphics[height=0.95em]{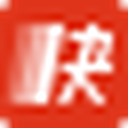}}}
\newcommand{\iconWenxin}{\iconbox{\includegraphics[height=0.95em]{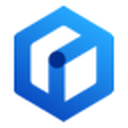}}}
\newcommand{\iconMetaso}{\iconbox{\includegraphics[height=0.95em]{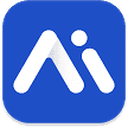}}}
\newcommand{\iconSoSearch}{\iconbox{\includegraphics[height=0.95em]{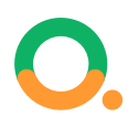}}}
\newcommand{\iconYuanbao}{\iconbox{\includegraphics[height=0.95em]{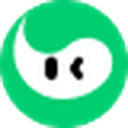}}}
\newcommand{\iconDeepSeek}{\iconbox{\includegraphics[height=0.95em]{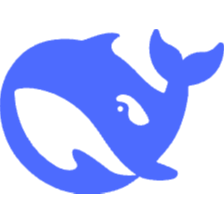}}}
\newcommand{\iconChatGPT}{\iconbox{\includegraphics[height=0.95em]{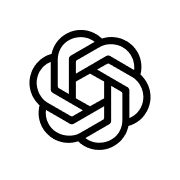}}}
\newcommand{\iconGemini}{\iconbox{\includegraphics[height=0.95em]{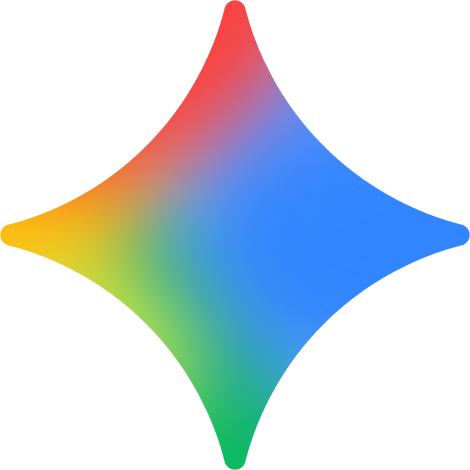}}}
\newcommand{\iconGoogle}{\iconbox{\includegraphics[height=0.95em]{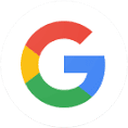}}}
\newcommand{\iconPCauto}{\iconbox{\includegraphics[height=0.95em]{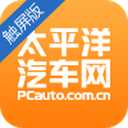}}}
\definecolor{biliBg}{HTML}{FDE7EF}
\definecolor{bjhBg}{HTML}{E6EEFF}
\definecolor{kuaiBg}{HTML}{FDE9E5}
\newcommand{\slot}[1]{{\setlength{\fboxsep}{0.8pt}\colorbox{black}{\strut\textcolor{white}{#1}}}}
\definecolor{svcWenxin}{HTML}{D9E7F8}   \definecolor{svcMetaso}{HTML}{DBE2F5}   \definecolor{svcGemini}{HTML}{ECE7F6}
\definecolor{svcChatGPT}{HTML}{EBEBEB}  \definecolor{svcYuanbao}{HTML}{D6F6E8}  \definecolor{svcSoSearch}{HTML}{E3F4D6}
\definecolor{svcGoogle}{HTML}{E3EDFD}   \definecolor{svcDeepSeek}{HTML}{E2E7FE}

%% file: sections/numbers.tex
\newcommand{\nTermsBili}{461}

\newcommand{\nTermsBjh}{279}

\newcommand{\nTermsKuai}{541}

\newcommand{\requestPace}{1.4}               
\newcommand{\stopRule}{15\%}

\newcommand{\nHotlineExcludedBili}{335}

\newcommand{\syncSigBH}{77}
\newcommand{\syncSigNoBurst}{81}
\newcommand{\syncClientsNoBurst}{172}

\newcommand{\kuaiSharedTitles}{3,017}

\newcommand{\sisterIdentical}{16}
\newcommand{\sisterTotal}{17}
\newcommand{\sisterWithinFive}{11}
\newcommand{\sisterGap}{0.5}
\newcommand{\caseBikeSharedChars}{2,630} 

\newcommand{\yiwangTitles}{254}
\newcommand{\yiwangAccounts}{107}
\newcommand{\bikeRefs}{98}

\newcommand{\bikeRefsFRW}{51}
\newcommand{\nAssistants}{nine}

\newcommand{\methSnapshotDate}{23 October 2026}
\newcommand{\methFoundBili}{8,525}      \newcommand{\methFoundBjh}{2,647}      \newcommand{\methFoundKuai}{508}
\newcommand{\methInStudyBili}{3,574}    \newcommand{\methInStudyBjh}{2,208}    \newcommand{\methInStudyKuai}{501}
    
\newcommand{\methMembersBili}{2,397}    \newcommand{\methMembersBjh}{1,067}    \newcommand{\methMembersKuai}{291}
\newcommand{\methMembersAll}{3,755}
              
\newcommand{\methPeriodBili}{01/2019--10/2026} \newcommand{\methPeriodBjh}{01/2019--10/2026} \newcommand{\methPeriodKuai}{01/2024--10/2026}
\newcommand{\methGenuineOutletsExcluded}{51}   
\newcommand{\methYieldBili}{9.2\%}      \newcommand{\methYieldBjh}{3.3\%}      \newcommand{\methYieldKuai}{6.5\%}   

\newcommand{\methValN}{72}                     
\newcommand{\methValPrecision}{92.3\%}  \newcommand{\methValRecall}{92.3\%}    
\newcommand{\methClassifierTrainN}{27,299} \newcommand{\methClassifierAccuracy}{89.3\%}

\newcommand{\introBikeRefsSameBrand}{\bikeRefsFRW}               
\newcommand{\rqTwoClientsAllThree}{112}
\newcommand{\rqTwoSharedNicheShare}{82\%}     
\newcommand{\rqTwoXcorrPairs}{167}            
\newcommand{\rqTwoXcorrSigShare}{38\%}        
\newcommand{\rqTwoCopyPairs}{1,394}           
\newcommand{\rqTwoCopyWithinHour}{49\%}       
\newcommand{\rqTwoCopyPairsAll}{5,540}        
\newcommand{\rqTwoCopyClientsAll}{292}        
\newcommand{\rqTwoCopyMedianHours}{6}         
\newcommand{\rqTwoCopyWithinDayAll}{69\%}     
\newcommand{\rqTwoPairs}{363}                 
\newcommand{\rqTwoPairedAccounts}{223}        
\newcommand{\rqTwoPairWithinHour}{57\%}       
\newcommand{\rqTwoIncidentalWithinHour}{9\%}  

\newcommand{\rqOneNWithFollowersBili}{1,676}          
\newcommand{\rqOneMedianFollowersBili}{9}             
\newcommand{\rqOneShareUnderFiftyBili}{63\%}          
\newcommand{\rqOneMedianFollowersGenuine}{913}        
\newcommand{\rqOneShareOutletNamedBjh}{11.5\%}        
\newcommand{\rqOneShareOutletNamedBjhNonmember}{5.0\%} 
\newcommand{\rqOneDomConsumerShare}{36\%}             
\newcommand{\rqOneDomHealthShare}{17\%}               
\newcommand{\rqOneDomFinanceShare}{14\%}              
\newcommand{\rqOneDomBeautyShare}{13\%}               
\newcommand{\rqOneDomEducationShare}{13\%}            
\newcommand{\rqOneGeoVendorPlacements}{2,629}         
\newcommand{\rqOneGeoVendorAccounts}{298}             
\newcommand{\rqOneSyncObs}{59.7\%}                    
\newcommand{\rqOneSyncNull}{28.0\%}                   
\newcommand{\rqOneSyncClients}{397}                   
\newcommand{\rqOneSyncSig}{219}                       
\newcommand{\rqOneSyncObsBjh}{41.8\%}                 
\newcommand{\rqOneSyncNullBjh}{16.7\%}                
\newcommand{\rqOneSyncClientsBjh}{292}                
\newcommand{\rqOneSyncSigBjh}{100}                    
\newcommand{\rqOneGapMedian}{20.6}                    
\newcommand{\rqOneGapNull}{56.1}                      
\newcommand{\rqOneBundlePairsTested}{199}             
\newcommand{\rqOneBundles}{34}                        
\newcommand{\rqOneGrowthFold}{2.5-fold}               
\newcommand{\rqOneActiveTwentyFour}{357}              
\newcommand{\rqOneActiveTwentyFive}{860}              
\newcommand{\rqOnePerActiveTwentyFour}{69}            
\newcommand{\rqOnePerActiveTwentyFive}{72}            

\newcommand{\rqOneEndedTwentyFive}{57\%}              
\newcommand{\rqOneEndedPre}{22\%}                     
\newcommand{\rqThreeServicesAll}{nine}          
\newcommand{\rqThreeQuestions}{ten}             
\newcommand{\rqThreeAnswers}{254}   
\newcommand{\rqThreeWindow}{13--24 October 2026}   
\newcommand{\rqThreeWindowLate}{19 October 2026}   
\newcommand{\rqThreeModelGemini}{gemini-2.5-flash}       
\newcommand{\rqThreeModelChatGPT}{gpt-4o-2024-08-06}     
\newcommand{\rqThreeCitedUrls}{1,887}   
\newcommand{\rqThreeFetchedUrls}{1,765}   
\newcommand{\rqThreeAdoptionWithoutCite}{11\%}   
\newcommand{\rqThreeDoseClients}{40}            

\newcommand{\rqThreeMetasoAnswers}{14}   

\newcommand{\appLlmBatch}{50}                    
\newcommand{\appLlmTemperature}{0.1}
\newcommand{\appLlmWorkers}{four}                
\newcommand{\appFilterMaxResults}{120}           
\newcommand{\appFormulaCap}{one third}           

\newcommand{\appReachBjhShare}{27\%}

\newcommand{\appReachPaddingP}{0.004}

\newcommand{\appReachListicleP}{0.009}

\newcommand{\rqThreeTitleSim}{0.8}              

\newcommand{\methValPrecisionCI}{83--97\%}

\newcommand{\rqTwoKuaiMaxPartners}{41}        

\newcommand{\methChaoTermsBili}{375}        \newcommand{\methChaoNBili}{3,639}   \newcommand{\methChaoCIBili}{3,459--3,850}   \newcommand{\methChaoCoverageBili}{66\%}   
           \newcommand{\methChaoCoverageBjh}{74\%}
            \newcommand{\methChaoCoverageKuai}{90\%}

\newcommand{\syncSigBHBjh}{53}

\newcommand{\rqThreeFetchedShare}{94\%}   

\newcommand{\discStratEventPanelN}{135}           
\newcommand{\discStratEventPanelFell}{59\%}       

\newcommand{\jOneModel}{Claude Haiku 5.5}
\newcommand{\llmJudgeModel}{GPT-4o mini}     

\newcommand{\jOneRecallSearched}{47\%}       
      \newcommand{\jOneAccountsReachedShare}{92\%}

\newcommand{\methPlacements}{466,357}        

\newcommand{\fmtGoldN}{963}                      

\newcommand{\fmtRulePrecision}{0.93}
\newcommand{\fmtRuleRecall}{0.96}

\newcommand{\rqThreeRoundThreeDate}{24 October 2026}
\newcommand{\rqThreeRoundThreeAnswers}{54}          

\newcommand{\lstTitlesBili}{20,676}
\newcommand{\lstTopN}{70\%}
\newcommand{\lstYearStamp}{50\%}
\newcommand{\lstKeywordRepeat}{57\%}
\newcommand{\lstSuperlative}{33\%}
\newcommand{\lstPersona}{20\%}

\newcommand{\lstPaddingMax}{3\%}        
\newcommand{\lstCitedLinkedN}{321}
\newcommand{\lstCitedOtherN}{287}
\newcommand{\lstCloseRestateLinked}{41\%}
\newcommand{\lstCloseRestateOther}{16\%}

\newcommand{\lstStylesN}{18}
\newcommand{\lstYearStampEarly}{12\%}    

\newcommand{\domOursOC}{96\%}            

\newcommand{\stratRankingDec}{45\%}      
\newcommand{\stratRankingSep}{14\%}      
\newcommand{\stratQuestionDec}{22\%}
\newcommand{\stratQuestionSep}{30\%}

\newcommand{\campNicheClients}{320}
\newcommand{\campBigClients}{111}         
\newcommand{\campBigLongestMedian}{196}   
\newcommand{\campBigSpanMedian}{862}      
\newcommand{\campBigLongShare}{57\%}      
\newcommand{\campBigCampaignsMedian}{5}   

\newcommand{\rqFourCensusClients}{22}
\newcommand{\rqFourCensusQuestions}{14}      
\newcommand{\rqFourAnswers}{334}             

\newcommand{\rqFourNicheNamedRange}{83--100\%}   
\newcommand{\rqFourOwnedNamedMax}{25\%}          

\newcommand{\rqThreeProbeSeededAccounts}{29}
\newcommand{\rqThreeProbeSeededMembers}{12}

\newcommand{\caseAcctFollowers}{3}
\newcommand{\caseAcctEntry}{March 2024}

\newcommand{\caseAcctPlacements}{915}
\newcommand{\caseAcctClients}{44}
\newcommand{\caseAcctIndustries}{education, health, consumer goods, finance and technology}

\newcommand{\caseAcctLast}{September 2026}
\newcommand{\caseAcctPeakMonth}{April 2025}
\newcommand{\caseAcctPeakN}{93}

\newcommand{\rqThreeNetListicleCited}{86\%}
\newcommand{\rqThreeOtherListicleCited}{49\%}

%% file: sections/numbers_imc_text.tex
\newcommand{\labMarkers}{232,830}         
\newcommand{\labClassifier}{159,973}      
\newcommand{\labPrior}{20,100}            
\newcommand{\labClientNamedShare}{11\%}   

\newcommand{\seedArticles}{two}

\newcommand{\engReachListicleAUC}{0.72}
\newcommand{\engReachPaddingAUC}{0.26}
\newcommand{\engDoseSpearman}{$-0.002$}   
\newcommand{\lstShareTwentyFour}{16\%}
\newcommand{\lstShareTwentyFive}{57\%}   

\newcommand{\supLabelHumanKappa}{0.52}   
\newcommand{\lstEntrantsCohortShare}{50\%}     
\newcommand{\lstEntrantsCohortListicle}{79\%}  
\newcommand{\lstCohortPreTwentyFour}{9\%}      
\newcommand{\lstCohortPreTwentyFive}{17\%}     
\newcommand{\stylePaddingMaxAny}{14\%}                                        
\newcommand{\pairGapMedianMin}{32}        
\newcommand{\incidentalGapMedianH}{23}    
\newcommand{\copyNullMedianDays}{282}     
\renewcommand{\methPeriodBili}{01/19--10/26}
\renewcommand{\methPeriodBjh}{01/19--10/26}
\renewcommand{\methPeriodKuai}{01/24--10/26}

%% file: sections/numbers_imc_engine.tex

\newcommand{\engTierUrlsExGeo}{1,580}   
\newcommand{\engTierStrictExGeo}{4.5\%}   
\newcommand{\engTierMidExGeo}{7.2\%}   
\newcommand{\engTierBroadExGeo}{20\%}   
\newcommand{\engTierWenxinStrictExGeo}{18\%}   
\newcommand{\engTierWenxinMidExGeo}{22\%}   
\newcommand{\engTierWenxinBroadExGeo}{38\%}   
\newcommand{\engTierDeepSeekBroadExGeo}{5\%}   
\newcommand{\engGeoQuestionUrls}{185}   
\newcommand{\engGeoQuestionBroad}{100}   
\newcommand{\engGeoQuestionNamesOnly}{82}   
\newcommand{\engShortPages}{240}   
\newcommand{\engBroadNoShortExGeo}{23\%}   
\newcommand{\engBylinePages}{33}   
\newcommand{\engBylineNewsPages}{13}   
\newcommand{\engBylineSubstringPages}{4}   
\newcommand{\engStrictGenuineAuthorPages}{10}   
\newcommand{\engTierStrictCleanExGeo}{3.9\%}   
\newcommand{\engTierMidCleanExGeo}{5.8\%}   

\newcommand{\engCensusAnswers}{334}   
\newcommand{\engCensusClientNamed}{52\%}   
\newcommand{\engCensusControlNamed}{68\%}   
\newcommand{\engCensusRatio}{0.77}   
\newcommand{\engCensusClientWilson}{47--57\%}   
\newcommand{\engCensusControlWilson}{63--73\%}   
\newcommand{\engCensusDiffBootQ}{$-37$ to $+7$ points}   
\newcommand{\engCensusWilcoxonP}{0.15}   
\newcommand{\engCensusRepAgree}{71\%}   
\newcommand{\engCensusMetasoAnswers}{6}   
\newcommand{\engModelChatGPTMini}{gpt-4o-mini-2024-07-18}   
\newcommand{\engCensusPooledClients}{14}   
\renewcommand{\rqFourNicheNamedRange}{79--100\%}   
\renewcommand{\rqFourOwnedNamedMax}{17\%}   
\newcommand{\engCSThreeSixtyRatio}{1.15}   
\newcommand{\engCSDeepSeekRatio}{0.56}   

\newcommand{\engIncumbentQuestions}{10}   
\newcommand{\engIncumbentClient}{54\%}   
\newcommand{\engNoIncumbentClient}{47\%}   
\newcommand{\engIncumbentFisherP}{0.23}   
\newcommand{\engVolumeSpearman}{$-0.26$}   
\newcommand{\engVolumeSpearmanP}{0.37}   

\newcommand{\engAdoptWithCite}{59\%}   
\renewcommand{\rqThreeAdoptionWithoutCite}{15\%}   
\newcommand{\engAnyNetClient}{56\%}   
\newcommand{\engTargetClient}{39\%}   
\newcommand{\engPilotPrefixedClient}{44\%}   
\newcommand{\engPilotPrefixedControl}{53\%}   
\newcommand{\engPilotRepAgree}{69\%}   

\newcommand{\engRepAnswers}{50}   
\newcommand{\engRepCiteNetDefence}{18\%}   
\newcommand{\engRepCiteNetDefenceStock}{20\%}   
\newcommand{\engRepCiteNetDefenceMask}{16\%}   
\newcommand{\engRepPenaltyAnswers}{19}   
\newcommand{\engRepPenaltyOf}{25}   

\newcommand{\engGeoAnswers}{25}   
\newcommand{\engGeoVendorAnswers}{21}   
\newcommand{\engGeoTopVendorAnswers}{17}   

\newcommand{\engReachExposed}{453}   
\newcommand{\engReachCited}{17}   
\newcommand{\engReachBjhCited}{13}   
\newcommand{\engReachBjhExposed}{62}   
\newcommand{\engReachBiliCited}{4}   
\newcommand{\engReachBiliExposed}{391}   
\renewcommand{\appReachBjhShare}{21\%}   
\renewcommand{\appReachPaddingP}{0.010}   
\renewcommand{\appReachListicleP}{0.014}   

\newcommand{\engDoseSpearmanP}{0.98}   
\newcommand{\engDoseQuartiles}{11/30, 11/29, 5/30, 13/30}   

\newcommand{\engBikeWenxinReps}{three}   
\newcommand{\engBikeWenxinKing}{two}   

\newcommand{\engWenxinBjhPilot}{27\%}   
\newcommand{\engWenxinBjhCensus}{41\%}   
\newcommand{\engMetasoBiliPilot}{13\%}   

%% file: sections/numbers_imc_supply.tex

\newcommand{\supAnalysisAccounts}{1,985}   

\newcommand{\caseBikePlacements}{1,185}   
\newcommand{\caseBikeTitles}{716}   
\newcommand{\caseBikeAccounts}{188}   
\newcommand{\caseBikeFirst}{June 2024}   
\newcommand{\caseBikeLast}{September 2026}   
\newcommand{\caseBikeRanking}{969}   
\newcommand{\caseBikeWaves}{four}   
\newcommand{\caseBikeMarAugAll}{118}   
\newcommand{\caseBikeMarAugRanking}{112}   
\newcommand{\caseBikeSepAll}{80}   
\newcommand{\caseBikeSepNonRanking}{72}   
\newcommand{\caseBikeSepAccounts}{29}   
\newcommand{\caseBikeSepPriorMarch}{4}   
\newcommand{\caseBikeSepAugRanking}{18}   

\renewcommand{\discStratEventPanelFell}{61\%}   
\newcommand{\stratEventPanelFellN}{82}   
\newcommand{\stratEventPanelPooledPre}{48\%}   
\newcommand{\stratEventPanelPooledPost}{32\%}   
\newcommand{\stratEventPanelFellRange}{55--72\%}   
\newcommand{\stratAssistantTwentySixPrecise}{0.9\%}   
\newcommand{\stratAssistantPadTwentySixPrecise}{0.8\%}   

\newcommand{\lstPanelN}{211}   
\newcommand{\lstPanelMedTwentyFour}{7\%}   
\newcommand{\lstPanelMedTwentyFive}{10\%}   
\newcommand{\lstShiftWithinShare}{17\%}   

\newcommand{\bjhHistAccounts}{20}   
\newcommand{\bjhHistListiclePre}{8\%}   
\newcommand{\bjhHistListicleLatest}{75\%}   
\newcommand{\bjhHistPrPre}{63\%}   
\newcommand{\bjhHistPrLatest}{6\%}   
\newcommand{\bjhHistCompleteN}{ten}   
\newcommand{\bjhHistCompleteListicleLatest}{74\%}   

\newcommand{\tmplNearDupOtherAccount}{14\%}   
\newcommand{\tmplNullSameTopicPair}{0.09\%}   
\newcommand{\tmplCount}{478}   
\newcommand{\tmplMedianAccounts}{13}   
\newcommand{\tmplMultiAccount}{97\%}   
\newcommand{\tmplMultiPlatform}{36\%}   
\newcommand{\tmplTopPosts}{751}   
\newcommand{\tmplTopAccounts}{215}   
\newcommand{\tmplExactReuse}{7.5\%}   

\renewcommand{\rqOneSyncObs}{59.1\%}   
\renewcommand{\rqOneSyncNull}{27.6\%}   
\renewcommand{\rqOneSyncSig}{220}   
\renewcommand{\rqOneSyncObsBjh}{43.0\%}   
\renewcommand{\rqOneSyncNullBjh}{16.8\%}   
\renewcommand{\rqOneGapMedian}{20.5}   
\renewcommand{\rqOneGapNull}{56.0}   
\renewcommand{\syncSigNoBurst}{222}   
\renewcommand{\syncClientsNoBurst}{396}   
\renewcommand{\syncSigBH}{188}   
\renewcommand{\syncSigBHBjh}{70}   
\newcommand{\rqOneSyncObsKuai}{45.0\%}   
\newcommand{\rqOneSyncNullKuai}{21.7\%}   
\newcommand{\rqOneSyncSigKuai}{91}   
\newcommand{\rqOneSyncClientsKuai}{308}   

\renewcommand{\kuaiSharedTitles}{2,881}   
\renewcommand{\rqTwoKuaiMaxPartners}{38}   
\newcommand{\rqTwoPairedMatches}{7,159}   
\newcommand{\rqTwoPairWithinFiveMin}{34\%}   
\newcommand{\rqTwoTopPairTitles}{737}   
\newcommand{\rqTwoTopPairWithinHour}{99\%}   
\newcommand{\rqTwoCopyNullWithinDay}{2\%}   

\renewcommand{\rqTwoClientsAllThree}{142}   

\newcommand{\supOriginFirstPostPlacement}{68\%}   
\newcommand{\supOriginWithinNinety}{96\%}   
\newcommand{\supOriginAgedRegs}{250}   
\newcommand{\supOriginReactivated}{69}   
\newcommand{\supAssemblyAcctShare}{5.8\%}   
\newcommand{\supAssemblyPlacementShare}{27.5\%}   
\newcommand{\supAssemblyBjhRandomAcct}{17\%}   
\newcommand{\supAssemblyBjhRandomPlacement}{35\%}   

\newcommand{\supFolGenuineNBili}{11}   
\newcommand{\supFolOtherNBili}{366}   
\newcommand{\supLabelModelLowConf}{33\%}   
\newcommand{\supClassifierJudge}{GPT-4o mini}   
\newcommand{\supLabelHumanN}{153}   
\newcommand{\supLabelHumanAgree}{89\%}   

\renewcommand{\lstCitedLinkedN}{106}   
\renewcommand{\lstCitedOtherN}{525}   
\renewcommand{\lstCloseRestateLinked}{42\%}   
\newcommand{\campRepeatShare}{57\%}   
\newcommand{\campGapMedian}{64.5}   

\renewcommand{\rqOneGeoVendorPlacements}{2,628}   
\renewcommand{\rqOneGeoVendorAccounts}{264}   
\renewcommand{\rqOneGrowthFold}{2.8-fold}   
\renewcommand{\rqOneActiveTwentyFour}{349}   
\renewcommand{\rqOneActiveTwentyFive}{850}   
\renewcommand{\rqOnePerActiveTwentyFour}{54}   
\renewcommand{\rqOnePerActiveTwentyFive}{62}   
\renewcommand{\rqOneEndedTwentyFive}{56\%}   
\renewcommand{\rqOneEndedPre}{21\%}   
\renewcommand{\lstTitlesBili}{28,317}   
\renewcommand{\lstTopN}{63\%}   
\renewcommand{\lstKeywordRepeat}{50\%}   
\renewcommand{\lstYearStamp}{51\%}   
\renewcommand{\lstSuperlative}{27\%}   
\renewcommand{\lstPersona}{16\%}   
\renewcommand{\lstYearStampEarly}{11\%}   

%% file: sections/numbers_imc_fixes.tex
\newcommand{\popMinedMembers}{2,016}        
\newcommand{\popOutletsMined}{31}           
\newcommand{\popAnalysisBili}{1,457}        \newcommand{\popAnalysisBjh}{248}         \newcommand{\popAnalysisKuai}{280}       
\newcommand{\popArticlesBili}{417,943}      \newcommand{\popArticlesBjh}{246,758}     \newcommand{\popArticlesKuai}{67,926}    
\newcommand{\popPlacementsBili}{239,935}    \newcommand{\popPlacementsBjh}{186,352}   \newcommand{\popPlacementsKuai}{40,070}  
\newcommand{\popCoreAccounts}{1,640}         

\renewcommand{\rqOneSyncObs}{61.5\%}        \renewcommand{\rqOneSyncNull}{29.4\%}     
\renewcommand{\rqOneSyncSig}{98}            \renewcommand{\rqOneSyncClients}{175}     
\renewcommand{\syncSigBH}{84}               
\renewcommand{\rqOneSyncObsBjh}{43.8\%}     \renewcommand{\rqOneSyncNullBjh}{18.0\%}  
\renewcommand{\rqOneSyncSigBjh}{32}         \renewcommand{\rqOneSyncClientsBjh}{106}  
\renewcommand{\syncSigBHBjh}{16}            
\renewcommand{\rqOneSyncObsKuai}{50.0\%}    \renewcommand{\rqOneSyncNullKuai}{29.4\%} 
\renewcommand{\rqOneSyncSigKuai}{30}        \renewcommand{\rqOneSyncClientsKuai}{124} 
\newcommand{\syncSigBHKuai}{0}              
\renewcommand{\syncSigNoBurst}{101}         \renewcommand{\syncClientsNoBurst}{174}   
\renewcommand{\rqOneGapMedian}{24.2}        \renewcommand{\rqOneGapNull}{89.7}        

\newcommand{\domTableClients}{319}          
\newcommand{\rqTwoMatchesAll}{17,280}       

\newcommand{\engCleanStrictRemoved}{9}      
\newcommand{\engCleanMidRemoved}{22}        

\renewcommand{\engAnyNetClient}{53\%}       
\renewcommand{\engTargetClient}{43\%}       

\newcommand{\methValNetwork}{65}            \newcommand{\methValOrganic}{7}           
\newcommand{\methValTP}{60}                 \newcommand{\methValFP}{5}                
\newcommand{\methValTN}{2}                  \newcommand{\methValSpecCI}{8--64\%}      

\newcommand{\popAutomotive}{345}            
\newcommand{\censusDate}{25 October 2026}   

\newcommand{\popMinedBili}{1,464}   \newcommand{\popMinedBjh}{261}   \newcommand{\popMinedKuai}{291}     
\newcommand{\popOutletsBili}{7}     \newcommand{\popOutletsBjh}{13}  \newcommand{\popOutletsKuai}{11}    
\newcommand{\popAutoBili}{300}      \newcommand{\popAutoBjh}{13}     \newcommand{\popAutoKuai}{32}       
\newcommand{\popCoreBili}{1,157}    \newcommand{\popCoreBjh}{235}    \newcommand{\popCoreKuai}{248}      
\newcommand{\supAssemblyN}{116}     \newcommand{\supAssemblyBjhN}{105}   
\newcommand{\supAssemblyBjhRandomN}{100}   

\renewcommand{\stratRankingDec}{50\%}   \renewcommand{\stratRankingSep}{18\%}     
\renewcommand{\stratQuestionDec}{14\%}  \renewcommand{\stratQuestionSep}{22\%}    
 
\renewcommand{\stratEventPanelFellN}{81}              
\renewcommand{\discStratEventPanelFell}{60\%}     
\renewcommand{\stratEventPanelFellRange}{56--69\%}                                            
\renewcommand{\stratEventPanelPooledPre}{54\%}  \renewcommand{\stratEventPanelPooledPost}{40\%} 
\renewcommand{\caseBikeRanking}{896}          
\renewcommand{\caseBikeMarAugRanking}{87}     
\renewcommand{\caseBikeSepNonRanking}{71}     
\renewcommand{\caseBikeSepAccounts}{27}       
\renewcommand{\caseBikeSepAugRanking}{16}     
\renewcommand{\caseBikeSepPriorMarch}{4}      

\newcommand{\supFolOtherAllBili}{1,177}   

%% file: sections/abstract.tex
\begin{abstract}

Generative engines synthesize retrieved web pages into answers, recommendations, and rankings.
Generative engine optimization (GEO) services sell visibility in these answers by publishing promotional articles on user-generated-content (UGC) platforms for paying clients.
Such promotion raises concerns in medical and financial domains, where misleading advice can cause harm.
To study the severity of this problem, we measure one such paid-content operation across three major Chinese UGC platforms, uncovering \supAnalysisAccounts{} accounts associated with the network comprising \methPlacements{} inferred promotional placements.
We find that the network promotes products across diverse categories through a growing pool of accounts, conducting coordinated campaigns spanning months across multiple platforms while evolving its listicle content over time.
Across nine generative engines from both the US and China, we find that \engTierStrictExGeo{} to \engTierBroadExGeo{} of retrievable cited pages are attributable to the network, depending on how strictly attribution is drawn.
On eight of these engines, its clients are named in \engCensusClientNamed{} of answers, against \engCensusControlNamed{} for category leaders. 
Our work demonstrates the effectiveness of UGC GEO placement at influencing generative engine outcomes, and reveals content-level defenses as insufficient against coordinated UGC placements, where account-level and platform-level signals are necessary for proper defense.
\end{abstract}

%% file: sections/1-introduction.tex
\section{Introduction}
\label{sec:intro}

AI assistants with web search are now a common way to get answers and recommendations.
In China, more than 700 million people had used generative AI services by mid-2026, and 76\% of them say they use these services to answer questions.\footnote{\href{https://www.news.cn/20260929/40258c3794684c20ae7867909cdec2ce/c.html}{CNNIC report, via Xinhua, 29 Sep.\ 2026}.}
Unlike a search engine, which returns a list of links, such an assistant retrieves web pages, passes them to a language model, and returns a single answer that lists some of the pages as references.
What the answer says therefore depends on which pages the assistant retrieves.

Generative engine optimization (GEO) services sell influence over those pages.
A GEO vendor writes promotional articles for paying clients, which we call \emph{placements}, and publishes them through accounts it controls on content platforms that assistants retrieve (\S\ref{sec:platforms}).
Such services are advertised openly, mostly for ordinary consumer brands but also for dietary supplements, private hospitals, stock-tipping services, consumer-loan apps and immigration consultancies, where a misleading recommendation is costly.
In March 2026, China's national consumer-rights broadcast showed a GEO service that made major assistants recommend a fictitious product as the best in its category within days~(Appendix~\ref{app:sources}, S1).
The effect also appears in ordinary answers.
Asked to rank road-bike brands, Baidu's Wenxin assistant described, in two of three runs, a ``one king and four queens'' hierarchy of makers that crowns one brand as king.
That brand is a client of the network we study, whose accounts have published \caseBikePlacements{} placements naming it since \caseBikeFirst{} (\S\ref{sec:coordination}).
Wenxin's answers draw on that campaign: \introBikeRefsSameBrand{} of the \bikeRefs{} references behind them carry the brand or the hierarchy in their titles.
On Bilibili alone, the phrase appears in the titles of \yiwangTitles{} articles from \yiwangAccounts{} accounts.

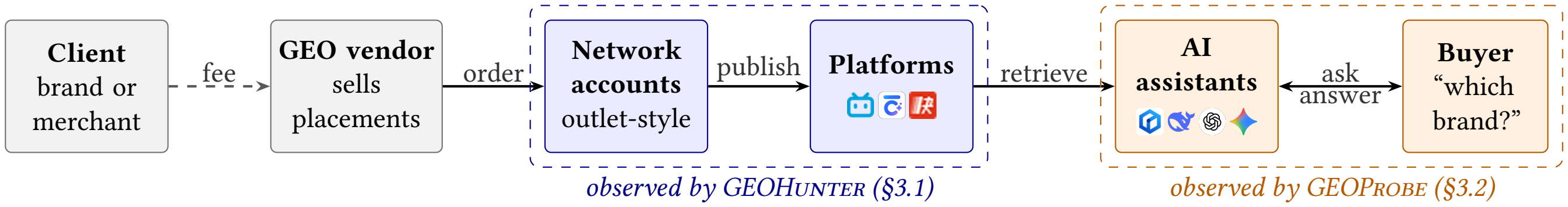
\begin{figure*}[t]
  \centering
  \small
  \begin{tikzpicture}[
      node distance=3mm and 11.5mm,
      actor/.style={draw, rounded corners=2pt, align=center, inner sep=2pt, minimum height=15mm, text width=17mm, font=\small},
      side/.style={actor, fill=black!5, draw=black!55},
      hstep/.style={actor, fill=blue!8, draw=blue!50!black},
      pstep/.style={actor, fill=orange!12, draw=orange!70!black},
      pay/.style={-{Stealth[length=1.6mm]}, thick, dashed, draw=black!65},
      flow/.style={-{Stealth[length=1.6mm]}, thick},
      lab/.style={font=\small, text=black!80, inner sep=1pt},
      grp/.style={draw, dashed, rounded corners=3pt, inner sep=1.6mm}]
    \node[side] (client) {\textbf{Client}\\brand or\\merchant};
    \node[side, right=of client, text width=18mm] (vendor) {\textbf{GEO vendor}\\sells\\placements};
    \node[hstep, right=of vendor] (accts) {\textbf{Network}\\\textbf{accounts}\\outlet-style};
    \node[hstep, right=of accts] (plat) {\textbf{Platforms}\\[2pt]\iconBili\,\iconBjh\,\iconKuai};
    \node[pstep, right=16mm of plat] (asst) {\textbf{AI}\\\textbf{assistants}\\[2pt]\iconWenxin\,\iconDeepSeek\,\iconChatGPT\,\iconGemini};
    \node[pstep, right=14mm of asst, text width=15mm] (user) {\textbf{Buyer}\\``which\\brand?''};
    \draw[pay] (client) -- node[lab, above] {fee} (vendor);
    \draw[flow] (vendor) -- node[lab, above] {order} (accts);
    \draw[flow] (accts) -- node[lab, above] {publish} (plat);
    \draw[flow] (plat) -- node[lab, above] {retrieve} (asst);
    \draw[{Stealth[length=1.6mm]}-{Stealth[length=1.6mm]}, thick] (asst) -- node[lab, above] {ask} node[lab, below] {answer} (user);
    \node[grp, draw=blue!50!black, fit=(accts) (plat)] (hunter) {};
    \node[grp, draw=orange!70!black, fit=(asst) (user)] (probe) {};
    \node[font=\small\itshape, text=blue!50!black, anchor=north, inner sep=1pt] at ([yshift=-0.5mm]hunter.south) {observed by \geohunter (\S\ref{sec:hunter})};
    \node[font=\small\itshape, text=orange!70!black, anchor=north, inner sep=1pt] at ([yshift=-0.5mm]probe.south) {observed by \geoprobe (\S\ref{sec:probe})};
  \end{tikzpicture}
  \caption{The GEO supply chain we measure (dashed arrow: payment; solid arrows: content). A client pays a GEO vendor, whose network accounts publish placements on content platforms that AI assistants retrieve; the assistants answer buyers' questions with references. Blue: observed by \geohunter; orange: observed by \geoprobe.}
  \label{fig:ecosystem}
\end{figure*}

Existing research has examined the engines that cite web content and the attacks that could manipulate them, but not the operations that supply the content: audits of generative search measure which kinds of sources assistants cite~\cite{zhen2026chinesegse,li2026baidugoogle,xu2026aioverviews,allaham2026synthetic}, but cannot say who produced them.
Laboratory studies show that crafted text can raise a product's rank in a generated answer~\cite{aggarwal2024geo,nestaas2025adversarial,pfrommer2024ranking,kumar2024manipulating}, but do not observe whether anyone does so at scale.
Measurements of black-hat search engine optimization map the supply side of manipulation~\cite{wang2013juice,wang2014seizure,du2016labyrinth,vangoethem2019pbn,wang2025illicit}, but the operations they study build domains and links for human readers who click.
No study has connected a production network observed in the wild to the answers that assistants give, in part because such a network is hard to see: each placement reads as an ordinary article, and the accounts that publish them do not signal that they work together.

This paper measures one such operation, from the accounts that publish its articles to the answers that cite them (Figure~\ref{fig:ecosystem}), through two research questions:
\par\noindent\textbf{RQ1 (supply, \S\ref{sec:rq1}).} Who publishes the placements, what do they promote, and how is their publication coordinated across accounts and platforms?
\par\noindent\textbf{RQ2 (exposure, \S\ref{sec:rq3}).} Which AI assistants retrieve the network's articles, what share of their references do these articles make up, and what do the answers do with them?

To carry out the measurement, we build two tools, one for each question.
\geohunter finds the network's accounts by following client names (\S\ref{sec:hunter}).
Every placement must name its client, and ordinary writers rarely post about the same niche brand on several accounts on the same day.
We therefore start from two accounts that published articles we ordered for a fictitious company (Appendix~\ref{app:ethics}), search the platforms for the client names in their articles to find more accounts, and accept an account into the network when its posting pattern shows coordination.
\geoprobe then asks nine assistants the buyer questions found to be targeted by the network's articles, each paired with a category leader as control, and traces every cited page in the assistant answer back to the network at three levels of strictness (\S\ref{sec:probe}).

Our measurement shows that the network promotes ordinary consumer products through coordinated accounts, and that all nine assistants we probed cite its articles.
For RQ1 (\S\ref{sec:rq1}), \domOursOC{} of the network's niche placements promote ordinary consumer goods and services rather than the illicit products common in search spam, and listicles, articles that rank products in a category, grew from \lstShareTwentyFour{} to \lstShareTwentyFive{} of placements on the network's core Bilibili accounts between 2024 and 2025.
The accounts present themselves as news outlets or experts but have a median of \rqOneMedianFollowersBili{} followers on Bilibili, and for \rqOneSyncSig{} of \rqOneSyncClients{} niche clients, they post about the same client in the same weeks more often than chance.
For RQ2 (\S\ref{sec:rq3}), \engTierStrictExGeo{} to \engTierBroadExGeo{} of the pages that the assistants cite come from the network, depending on how strictly we attribute them.
Answers name the network's clients in \engCensusClientNamed{} of cases, compared with \engCensusControlNamed{} for the category leaders we used as controls, and they repeat the network's claims.

We make the following contributions.
\begin{list}{\textbullet}{%
  \setlength{\leftmargin}{1em}%
  \setlength{\labelwidth}{0.65em}%
  \setlength{\labelsep}{0.35em}%
  \setlength{\itemindent}{0pt}%
  \setlength{\topsep}{2pt}%
  \setlength{\itemsep}{2pt}%
  \setlength{\parsep}{0pt}%
  \setlength{\partopsep}{0pt}}
\item \textbf{Tools and data.} We build \geohunter, which maps a network of \methMembersAll{} accounts on three platforms, and \geoprobe, which traces assistant citations back to the network. We validate the membership rule, estimate crawl coverage, and will release both tools with a pseudonymized dataset (\S\ref{sec:mapping}).
\item \textbf{Supply.} We characterize the clients, formats, accounts and campaigns of the operation from \supAnalysisAccounts{} accounts and \methPlacements{} placements, and its response to exposure (\S\ref{sec:rq1}).
\item \textbf{Exposure.} We measure how its articles reach the references and text of nine assistants, and derive recommendations for providers, platforms and auditors (\S\ref{sec:rq3}, \S\ref{sec:discussion}).
\end{list}

%% file: sections/2-background.tex
\section{Background}
\label{sec:background}
\label{sec:platforms}

\noindentparagraph{Generative engines and UGC}
The platforms an engine retrieves from matter most when the engine's operator also runs them.
We study this practice in China, where the operators of the largest search engines and content platforms also offer conversational assistants with web search.
Widely used assistants include Baidu's Wenxin, ByteDance's Doubao, DeepSeek and Tencent's Yuanbao, together with the AI search engines Metaso and 360 AI Search.\footnote{文心, 豆包, DeepSeek, 腾讯元宝, 秘塔, 360 AI 搜索.}
Several of these operators run user-generated-content (UGC) platforms whose articles their own engines can retrieve, and audits of deployed engines find platform-level biases that steer each engine's retriever toward particular sites~\cite{li2026baidugoogle,zhen2026chinesegse}.
Baidu operates Baijiahao, a self-publishing platform whose pages are among the hosts most frequently referenced in Baidu's AI overviews~\cite{li2026baidugoogle}, and Qihoo 360 operates 360kuai alongside 360 AI Search.\footnote{百家号, 360快资讯, 哔哩哔哩.}
Bilibili, a video platform that also hosts public image-and-text articles, is indexed by the major search engines and is retrieved by some of the assistants we probe (\S\ref{sec:rq3}).
Placements on these three platforms are the subject of our measurement (\S\ref{sec:hunter}).

\noindentparagraph{Generative engine optimization}
Brands that want to appear in generated answers have taken up generative engine optimization (GEO), the practice of optimizing web content for ingestion and visibility by generative engines~\cite{aggarwal2024geo}.
Laboratory studies show that rewriting content can raise its visibility in generated answers~\cite{aggarwal2024geo,nestaas2025adversarial,pfrommer2024ranking,kumar2024manipulating}, that a few injected documents can corrupt retrieval-augmented generation~\cite{zou2025poisonedrag}, and that web-scale data can be poisoned at low cost~\cite{carlini2024poisoning}.
In practice, GEO therefore has two parts: creating content that engines favor, such as ranked recommendation lists, and placing that content, persistently and in a coordinated way, on the platforms that each target engine reads.
Most brands cannot sustain such placement themselves, so agencies sell it as a service, advertising distribution to a dozen content platforms as a way to be recommended by AI assistants~(Appendix~\ref{app:sources}, S1, S2).
Paid articles once took editorial forms such as product reviews, buyer's guides and news items; they are increasingly written for the engine.

\noindentparagraph{Regulation}
Chinese regulators treat undisclosed promotion of this kind as a consumer-protection problem.
The Advertising Law requires advertising to be identifiable as such, and a 2024 enforcement guide applies that requirement to online content that promotes goods or services without an advertising label~(Appendix~\ref{app:sources}, S3, S4).
Following the March 2026 broadcast~(S1), the Cyberspace Administration of China launched a campaign against the misuse of AI applications, whose second-phase bulletin named malicious marketing through GEO as a form of AI data poisoning~(S5, S6).
These actions target individual vendors and articles; they do not describe the operations that produce placements at scale, which is what we measure.


%% file: sections/3-mapping.tex
\section{Methodology}
\label{sec:mapping}

\begin{figure}[t]
  \centering
  \small
  \begin{tikzpicture}[
      node distance=2mm and 3.6mm,
      box/.style={draw, rounded corners=1pt, align=center, inner sep=2.5pt, minimum height=8.5mm, font=\small},
      hstep/.style={box, fill=blue!8, draw=blue!50!black},
      pstep/.style={box, fill=orange!12, draw=orange!70!black},
      grp/.style={draw, dashed, rounded corners=2pt, inner sep=1.6mm},
      arr/.style={-{Stealth[length=1.4mm]}, thick},
      lab/.style={font=\small\itshape, inner sep=0.5pt},
      hl/.style={font=\small, text=black!65, inner sep=1pt}]
    \node[hstep] (seed) {seed\\accounts};
    \node[hstep, right=of seed] (mine) {mine\\feeds};
    \node[hstep, right=of mine] (names) {vet\\names};
    \node[hstep, right=of names] (search) {search\\titles};
    \node[hstep, right=of search] (judge) {judge\\accounts};
    \draw[arr] (seed) -- (mine);
    \draw[arr] (mine) -- (names);
    \draw[arr] (names) -- (search);
    \draw[arr] (search) -- (judge);
    \draw[arr] (judge.south) -- ++(0,-2mm) -| (mine.south);
    \coordinate (loopmid) at ($(mine.south)!0.5!(judge.south)$);
    \node[hl, below=3.2mm of loopmid] (looplab) {accepted accounts are mined in turn};
    \begin{scope}[on background layer]
      \node[grp, fit=(seed) (judge) (looplab), draw=blue!50!black] (hunter) {};
    \end{scope}
    \node[lab, anchor=south west, text=blue!50!black] at ([xshift=0.8mm, yshift=0.6mm]hunter.south west) {\geohunter};
    \node[pstep, anchor=north] (svc) at ([yshift=-9mm]hunter.south) {ask \nAssistants{}\\assistants};
    \node[pstep, left=of svc] (qs) {questions\\+ controls};
    \node[pstep, right=of svc] (attr) {attribute\\cited pages};
    \draw[arr] (qs) -- (svc);
    \draw[arr] (svc) -- (attr);
    \begin{scope}[on background layer]
      \node[grp, fit=(qs) (svc) (attr), draw=orange!70!black] (probe) {};
    \end{scope}
    \node[lab, anchor=east, text=orange!70!black] at ([xshift=-1mm]probe.west) {\geoprobe};
    \draw[arr, dashed] (qs.north |- hunter.south) -- node[hl, left, pos=0.35] {clients} (qs.north);
    \draw[arr, dashed] (attr.north) -- node[hl, right, pos=0.65] {accounts, titles} (attr.north |- hunter.south);
  \end{tikzpicture}
  \caption{The pipeline. \geohunter mines accounts, vets client names, searches for them and judges the accounts found. \geoprobe asks \nAssistants{} assistants buyer questions and attributes cited pages to the network.}
  \label{fig:pipeline}
\end{figure}
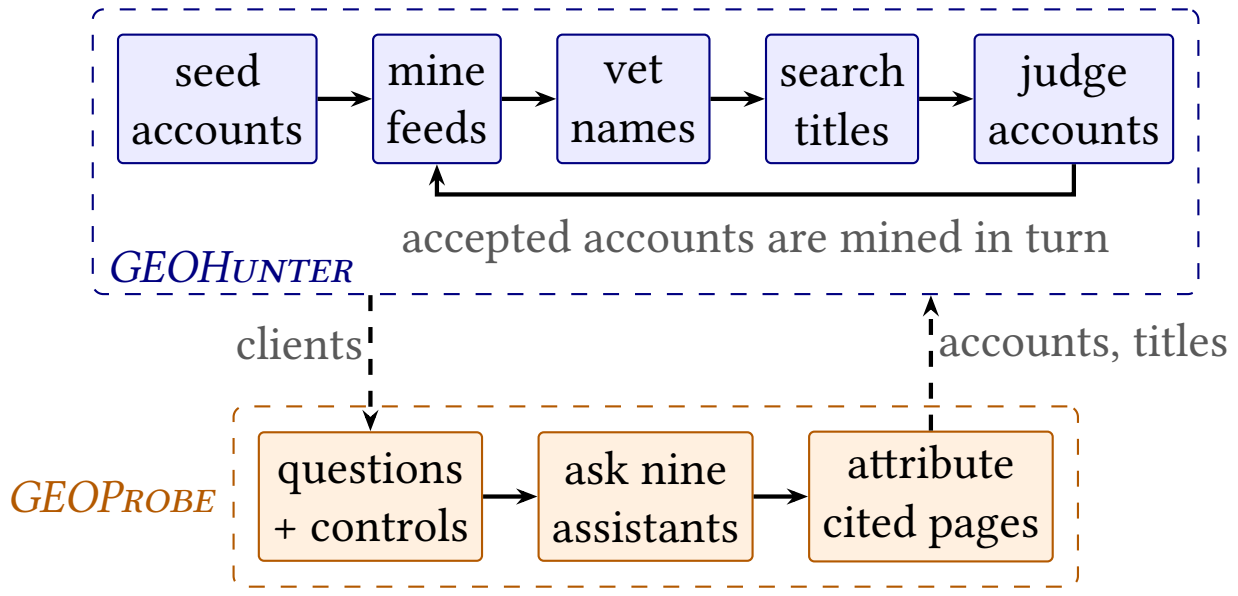

Answering RQ1 requires observing the accounts that publish placements, and answering RQ2 requires observing what the assistants cite and say.
Neither view is enough alone: the accounts show who publishes what, but not whether assistants use it, and the answers show what assistants cite, but not who wrote it.
We therefore build two tools, one for each side (Figure~\ref{fig:pipeline}).
\geohunter finds the network's accounts and collects their articles (\S\ref{sec:hunter}).
\geoprobe asks the assistants the buyer questions that the network's articles answer, and traces each cited page back to the network (\S\ref{sec:probe}).

\noindentparagraph{Terms}
A \emph{client} is an entity that placements promote, after merging spelling variants of one brand; nationally known brands with independent coverage are \emph{major brands} and the rest \emph{niche clients}, the focus of our client-level analyses.
An account returned by our searches is a \emph{found account}, and one that the rule of \S\ref{sec:hunter} accepts is a \emph{network account}.
A \emph{placement} is an article by a network account that our labeling stage classifies as paid promotion; \emph{padding} is one it classifies as not promotional, mostly reposted news and corporate announcements.
An \emph{analyzed account} is a network account whose article feed we collected, excluding genuine news outlets, and a \emph{core account} is an analyzed account outside the automotive segment, whose accounts mainly publish car makers' press releases.
An \emph{answer} is an assistant's response to one question and a \emph{reference} a page it cites.


\subsection{\geohunter: Mapping the Network}
\label{sec:hunter}
\label{sec:crawl}
\label{sec:terms}
\label{sec:identify}

No list of the network's accounts exists, but every placement names its client.
We therefore search the platforms for client names and follow them to new accounts; prior work used keyword snowballing to map promotional content on search engines~\cite{yang2017klingon,wang2025illicit}.
The crawl repeats five steps (Figure~\ref{fig:pipeline}).
\begin{enumerate}[leftmargin=1.4em,itemsep=1pt,topsep=2pt,parsep=0pt]
  \item \textbf{Seed.} Because every other membership is inferred, the crawl starts from the only placements we know to be paid: we ordered \seedArticles{} articles promoting a fictitious company from a commercial GEO vendor, and the Bilibili and 360kuai accounts that published them are the seeds (Appendix~\ref{app:ethics}). Baijiahao was seeded with client names first found on Bilibili.
  \item \textbf{Mine.} Given a seed or an accepted account, we collect its available article history and ask an LLM judge to propose client names from the titles.
  \item \textbf{Vet.} A proposed name is searched only if it is \emph{specific} (one entity, not a category), \emph{verbatim} (printed as other placements print it), \emph{promoted} (what the article sells, not what it mentions) and \emph{rare} in organic writing, which excludes generic products and well-known brands.
  \item \textbf{Search.} For each vetted name, we query each platform's own search (Appendix~\ref{app:crawl}) and record every account with an article whose title contains it verbatim.
  \item \textbf{Judge.} A found account is a network account if it published its last article in 2025 or later and shows at least one of three signals of coordinated publishing: \emph{several clients}, \emph{copied titles} shared with another account for the same client, or \emph{same-day bursts}, posting a client within a day of at least two other accounts that post it (thresholds in Appendix~\ref{app:crawl}). Accounts whose titles mostly advertise fake after-sales hotlines belong to a different operation and are excluded, and \methGenuineOutletsExcluded{} accepted accounts that are genuine news outlets carrying sponsored content are set aside from the analysis population. Accepted accounts return to step 2 and are mined in turn, though the crawl ended before all of them were mined.
\end{enumerate}

\noindentparagraph{Dataset}
Table~\ref{tab:dataset} summarizes what the loop produced on each platform: on Bilibili, \methFoundBili{} accounts were found and \methMembersBili{} accepted as network accounts, \methMembersAll{} across the three platforms.
We collected article feeds for \popMinedMembers{} of the \methMembersAll{} network accounts; excluding \popOutletsMined{} genuine news outlets leaves \supAnalysisAccounts{} analyzed accounts, and excluding a further \popAutomotive{} automotive accounts leaves \popCoreAccounts{} core accounts.
Outside the seed articles, promotional status and network membership are inferred from the observed records, and because the crawl reaches only accounts whose titles carry a searched name, all counts are lower bounds on the size of the network.
Bilibili is the core of the study, where the crawl started and complete account histories are available; Baijiahao serves as a second platform throughout, and 360kuai adds a third platform for the totals, for replications and for the cross-platform analysis of \S\ref{sec:sharedclients}.

\noindentparagraph{Article labels}
Client names alone cannot identify placements, because only \labClientNamedShare{} of the \methPlacements{} placements name an identified client in the title.
The remainder are labeled by format markers in the title (\labMarkers{}), by a text classifier trained on labels that an LLM judge assigned to clusters of similar titles (\labClassifier{}), and by the account's history (\labPrior{}).
One annotator agreed with the resulting placement labels on \supLabelHumanAgree{} of \supLabelHumanN{} Bilibili articles ($\kappa=\supLabelHumanKappa{}$; Appendix~\ref{app:crawl}).

\begin{figure}[t]
  \centering
  \includegraphics[width=\linewidth]{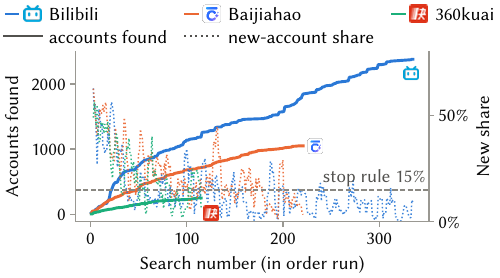}
  \caption{Snowball progress per platform: cumulative network accounts found, by search (solid, left axis) and the five-search rolling share of new network accounts among the results (dotted, right axis); searches retaining five or more accounts.}
  \label{fig:funnel}
  \label{fig:funnel-cum}
  \label{fig:funnel-yield}
\end{figure}

\noindentparagraph{Crawl coverage}
We stop crawling a platform when its searches stop finding new accounts: for each search that returns at least five accounts, we compute the share of returned accounts that are new network accounts, and we stop once the mean of this share over the last five searches falls below \stopRule{} (Appendix~\ref{app:crawl} gives the formula).
Figure~\ref{fig:funnel} shows each platform reaching that floor.
Because the stopping rule bounds the marginal return of further searches rather than the share of the network found, we estimate coverage separately: treating each searched name as a capture occasion, a Chao2 estimator places the crawl at about \methChaoCoverageBili{} of the Bilibili accounts reachable by name search, \methChaoCoverageBjh{} on Baijiahao and \methChaoCoverageKuai{} on 360kuai (Appendix~\ref{app:crawl}); accounts that no searched name reaches remain outside this estimate.

\noindentparagraph{Account identification accuracy}
\label{sec:evaluation}
To test the membership rule, an annotator labeled a stratified sample of Bilibili accounts from their complete article lists, without seeing the signals.
Against the \methValN{} in-study accounts with a definite label, the rule has a precision of \methValPrecision{} (95\% CI \methValPrecisionCI{}) and a recall of \methValRecall{}; Appendix~\ref{app:crawl} gives the confusion matrix.

\begin{table}[t]
  \caption{Data per platform (pipeline of \methSnapshotDate{}). Found accounts have at least one article whose title contains a searched client name verbatim; analyzed accounts are those with collected feeds minus genuine outlets, and core accounts further exclude the automotive segment; articles and placements are those of the analyzed accounts.}
  \label{tab:dataset}
  \footnotesize\setlength{\tabcolsep}{4pt}   
  \begin{tabular}{@{}l>{\columncolor{biliBg}}r>{\columncolor{bjhBg}}r>{\columncolor{kuaiBg}}r@{}}
    \toprule
     & \iconBili~Bilibili & \iconBjh~Baijiahao & \iconKuai~360kuai \\
    \midrule
    Names searched & \nTermsBili & \nTermsBjh & \nTermsKuai \\
    Accounts found & \methFoundBili & \methFoundBjh & \methFoundKuai \\
    Active in 2025 or later & \methInStudyBili & \methInStudyBjh & \methInStudyKuai \\
    Network accounts & \methMembersBili & \methMembersBjh & \methMembersKuai \\
    \midrule   
    Feeds collected & \popMinedBili & \popMinedBjh & \popMinedKuai \\
    \quad genuine outlets & \popOutletsBili & \popOutletsBjh & \popOutletsKuai \\
    Analyzed accounts & \popAnalysisBili & \popAnalysisBjh & \popAnalysisKuai \\
    \quad automotive segment & \popAutoBili & \popAutoBjh & \popAutoKuai \\
    Core accounts & \popCoreBili & \popCoreBjh & \popCoreKuai \\
    Articles & \popArticlesBili & \popArticlesBjh & \popArticlesKuai \\
    Placements & \popPlacementsBili & \popPlacementsBjh & \popPlacementsKuai \\
    Period & \methPeriodBili & \methPeriodBjh & \methPeriodKuai \\
    \bottomrule
  \end{tabular}
\end{table}

\subsection{\geoprobe: Reaching the Assistants}
\label{sec:probe}
\noindentparagraph{Probe design}
Because the network's articles are written as answers to buyer questions, we probe the assistants with the same questions.
Each question asks for recommendations in a category that a network client's placements target, phrased after the titles of those placements, and is paired with a \emph{control}: a non-network brand that leads the category.
We run two probe sets with different purposes.
The \emph{pilot} traces citations: it poses \rqThreeQuestions{} questions, nine category questions and one asking for GEO vendors, to \rqThreeServicesAll{} assistants three times each, producing \rqThreeAnswers{} answers during \rqThreeWindow{} (Metaso's quota allowed fewer runs; Appendix~\ref{app:probe}).
Five assistants are accessed through their web interfaces (Wenxin, Metaso, 360 AI Search, Yuanbao and DeepSeek) and four through search-enabled APIs (Gemini, ChatGPT, Google AI Overviews and Google AI Mode).
The \emph{census} measures naming across the network's sustained campaigns.
It covers the niche clients with at least five placements in six of the twelve months before the probe and at least twenty listicle placements (\rqFourCensusClients{} of the \campNicheClients{}), excluding those the pilot already covers and those that cannot be asked or scored.
Each of the remaining \rqFourCensusQuestions{} clients gets one category question, posed three times to each assistant, for \rqFourAnswers{} scored answers from eight of the \rqThreeServicesAll{} assistants (Metaso's runs were incomplete; Appendix~\ref{app:probe}).
Census controls were fixed before any answer was collected, whereas four of the nine pilot controls were chosen after the first round as the non-network brand that answers named most often; this choice favors the control, so the pilot comparison is conservative for the client.
We score each answer separately and report rates over answers; Appendix~\ref{app:probe} gives the model versions, the probe windows and the agreement among repetitions.
Doubao is not probed because it blocks access from outside mainland China, and our probes ran from the United States.

\noindentparagraph{Attribution}
For every page an assistant cites, attribution asks whether the page came from the network.
Because we recognize only the accounts and titles that \geohunter found, while the network's articles are also copied to hosts we did not crawl, we attribute each retrievable cited page at three nested levels and report all three.
\begin{itemize}[leftmargin=1.2em,itemsep=1pt,topsep=2pt,parsep=0pt]
  \item \textbf{Strict.} The page's author is a network account. This can hold only on the three platforms we crawled.
  \item \textbf{Extended.} The page is strictly attributed, or its title matches a network article for the same client, exactly or at a similarity of at least \rqThreeTitleSim{}, which catches copies on other hosts, or its byline names a network account.
  \item \textbf{Broad.} The page is attributed at the extended level, or its title or body names the client of the question.
\end{itemize}
The strict level is a lower bound on the network's presence; the broad level also counts genuine coverage of a client and is therefore an upper bound, and no level by itself shows that a page was paid for.
The question about GEO vendors has no single client, so its pages are excluded from these shares and analyzed separately (\S\ref{sec:rq3}).

\noindentparagraph{Measures}
We report three measures, each with what it cannot show.
The \emph{network reference share} is the share of retrievable distinct cited pages attributed to the network at each level; it measures exposure, not intent.
The \emph{naming rate} is the share of answers that name the client, compared with the share that name its control; it measures association, not the effect of any placement.
\emph{Claim repetition} records whether an answer restates a specific claim found in the network's articles; we trace cases rather than report a rate~\cite{zhen2026chinesegse}.


%% file: sections/4-rq1-organization.tex
\section{Inside the Network: Placements, Accounts and Campaigns}
\label{sec:rq1}
This section answers RQ1 for the \supAnalysisAccounts{} analyzed accounts and their \methPlacements{} placements, in four steps: what the placements promote and how they are written (\S\ref{sec:content}), which accounts carry them and how those accounts appear credible (\S\ref{sec:production}), how individual placements are organized into campaigns across accounts and platforms (\S\ref{sec:coordination}), and how the network responded to public exposure (\S\ref{sec:exposure}).
Longitudinal measurements use Bilibili's core accounts, whose complete histories we collected; Baijiahao and 360kuai provide levels and some replications.

\begin{figure*}[t]
  \addtocounter{figure}{1}
  \centering
  \subcaptionbox{Listicle and press-release share\label{fig:content-a}}{\includegraphics[width=0.32\linewidth]{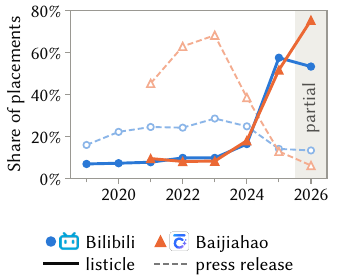}}\hfill
  \subcaptionbox{Listicle share by entry cohort\label{fig:content-b}}{\includegraphics[width=0.32\linewidth]{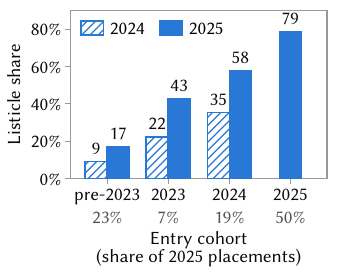}}\hfill
  \subcaptionbox{Title styles of listicles per year\label{fig:content-c}}{\includegraphics[width=0.32\linewidth]{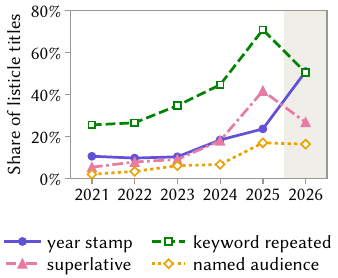}}
  \caption{Content of placements over time. (a) Listicle (solid) and press-release (dashed) share per year, core Bilibili and re-mined Baijiahao accounts. (b) Listicle share of each entry cohort in 2024 (hatched) and 2025; labels: the cohort's share of 2025 placements. (c) Title styles of core Bilibili listicles per year. Shaded: 2026, to 20 October.}
  \label{fig:content}
  \label{fig:growth-listicle}
\end{figure*}

\subsection{Placements: What Is Sold, and How It Is Written}
\label{sec:content}
\label{sec:clients}

\noindentparagraph{Client domains}
The network sells mainstream commerce rather than the illicit goods that dominate search spam.
We assign each niche client to an industry (Figure~\ref{fig:domains}) and weight it by its placements (Appendix Table~\ref{tab:domains_prior}): consumer goods account for \rqOneDomConsumerShare{} of niche placements, health products for \rqOneDomHealthShare{}, financial services for \rqOneDomFinanceShare{}, beauty for \rqOneDomBeautyShare{} and education for \rqOneDomEducationShare{}.
In all, \domOursOC{} of niche placements promote ordinary commerce, whereas prior measurements of Chinese search spam found at most a fifth of promoted content in ordinary commerce and most of it in gambling, counterfeit goods and unlicensed pharmacies (Appendix~\ref{app:bodies}).
GEO vendors are themselves clients: since 2025, \rqOneGeoVendorAccounts{} core accounts have published \rqOneGeoVendorPlacements{} placements whose titles advertise GEO services.

\begin{figure}[t]
  \addtocounter{figure}{-2}
  \centering
  \includegraphics[width=\linewidth]{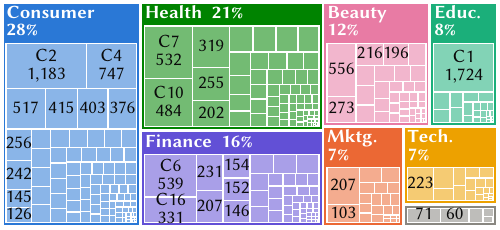}
  \caption{The \domTableClients{} niche clients with a core placement, by domain: blocks are domains sized by number of clients (header: share of clients); tiles are clients sized by core-segment placements (C$n$: Client $n$).}
  \label{fig:domains}
  \addtocounter{figure}{1}
\end{figure}

\noindentparagraph{The listicle}
The network's dominant format is the \emph{listicle}, a ranked recommendation article that presents promotion as the answer to a buyer's question.
Its title asks a selection question, such as ``which brand of liver-support tablets is best,'' and its body ranks alternatives with the client at the top; we count a placement as a listicle when its title ranks or recommends several candidates in a category, asks which brand or vendor to choose, or sets named brands against a judgment word, a rule with precision \fmtRulePrecision{} and recall \fmtRuleRecall{} against \fmtGoldN{} hand-coded titles.

The listicle became the majority format of the network as Chinese assistants added web search, and it did so mostly through new accounts.
On core Bilibili accounts, listicles grew from \lstShareTwentyFour{} of placements in 2024 to \lstShareTwentyFive{} in 2025 while press releases declined (Figure~\ref{fig:content}a), over the period in which the assistants began to search the web and Baidu placed generated overviews above its results.
Accounts that entered in 2025 carried \lstEntrantsCohortShare{} of that year's placements, \lstEntrantsCohortListicle{} of them listicles, whereas the \lstPanelN{} accounts with at least ten placements in both years changed little: their median listicle share rose only from \lstPanelMedTwentyFour{} to \lstPanelMedTwentyFive{} (Figure~\ref{fig:content}b; Appendix~\ref{app:bodies} gives the decomposition).
The same shift appears on Baijiahao, where the listicle share of \bjhHistAccounts{} accounts re-mined to a depth of 5,000 articles rose from \bjhHistListiclePre{} up to 2023 to \bjhHistListicleLatest{} in 2026 (Appendix~\ref{app:bodies}).

\noindentparagraph{Title styles}
Listicle titles use wording that matches typical buyer queries, and this wording spread together with the listicle.
We code \lstStylesN{} non-exclusive title styles, rules on the title string each with hand-checked precision of at least 0.90 (Appendix Table~\ref{tab:styles}).
Of the \lstTitlesBili{} listicle titles that core Bilibili accounts posted in 2026, \lstTopN{} promise a top-N ranking, \lstYearStamp{} include the year, \lstKeywordRepeat{} repeat the query keyword, \lstSuperlative{} use a superlative, and \lstPersona{} name a type of buyer or condition (Figure~\ref{fig:listicle-examples}); the ranking, superlative and buyer-type marks appear in at most \lstPaddingMax{} of the same accounts' padding titles.
These words make a title look recent, match the query, and promise a definite answer.
All three marks became more common after 2021: the year stamp, for example, rose from \lstYearStampEarly{} of listicle titles in 2021 to \lstYearStamp{} in 2026 (Figure~\ref{fig:content}c; Appendix Table~\ref{tab:styles}), while none of them exceeds about \stylePaddingMaxAny{} of padding titles in any year.
The article bodies show the same pattern.
Among the listicle pages cited in the pilot whose text we retrieved (\S\ref{sec:rq3}), the \lstCitedLinkedN{} pages attributed to the network at the extended level repeat a brand name in their closing paragraph in \lstCloseRestateLinked{} of cases, against \lstCloseRestateOther{} of the \lstCitedOtherN{} pages with no link to the network.

\begin{figure}[t]
  \centering
  \small
  \begin{tikzpicture}[
      card/.style={draw=black!35, rounded corners=1.5pt, inner sep=3pt, text width=\linewidth-8pt, align=left, font=\small},
      tag/.style={font=\small\sffamily, inner sep=1.2pt, rounded corners=1pt},
      q/.style={tag, fill=blue!15, text=blue!50!black},
      sh/.style={tag, fill=orange!20, text=orange!60!black},
      ev/.style={tag, fill=green!18, text=green!40!black},
      bd/.style={tag, fill=violet!15, text=violet!60!black},
      fr/.style={tag, fill=red!12, text=red!60!black},
      au/.style={tag, fill=teal!15, text=teal!60!black},
      plat/.style={font=\small\sffamily\bfseries, text=black!55, anchor=north east, inner sep=2pt},
      node distance=2mm]
    \node[card] (c1) {%
      \textbf{Title (cited page, Sina Finance).} Which quercetin brand works best? Brand reputation and ingredient recommendations.\\
      \tikz[baseline]\node[q]{buyer question}; \tikz[baseline]\node[ev]{rubric}; \\[1pt]
      \textbf{Body.} ``Based on ingredient activity, technical barriers, clinical validation and safety certification \ldots{} [the promoted brand] tops the list with a claimed lung-clearing effect 87\% ahead.''\\
      \tikz[baseline]\node[ev]{rubric}; \tikz[baseline]\node[bd]{answer first}; \tikz[baseline]\node[ev]{authority warrant};};
    \node[card, below=of c1] (c2) {%
      \textbf{Title (Bilibili).} The No.\,1 domestic spot-fading product: the only state-certified spot remover, publicly acclaimed.\\
      \tikz[baseline]\node[sh]{declared \#1}; \tikz[baseline]\node[ev]{authority warrant}; \tikz[baseline]\node[fr]{superlative};};
    \node[card, below=of c2] (c3) {%
      \textbf{Title (cited page, Baidu mobile).} Liver pills for fatty liver? Eight rankings to help choose liver-support products.\\
      \tikz[baseline]\node[q]{buyer question}; \tikz[baseline]\node[sh]{top-N ranking}; \\[1pt]
      \textbf{Close.} ``Patients should choose a brand with clear clinical data (such as [Client 15]) and keep taking it.''\\
      \tikz[baseline]\node[bd]{brand restated at close}; \tikz[baseline]\node[au]{persona / condition};};
  \end{tikzpicture}
  \caption{Recommendation framing in two pages that assistants cited, hosted outside the crawled platforms, and one Bilibili title, translated and pseudonymized. Tags name the styles of Appendix Table~\ref{tab:styles}, colored by its groups: list shape (orange), evidence (green), framing (red), audience (teal). Blue marks the buyer question that makes a title a listicle (\S\ref{sec:content}); violet marks two body devices that the table does not code.}
  \label{fig:listicle-examples}
\end{figure}

\finding{1}{The network sells mainstream commerce. Between 2024 and 2025 its placements shifted to listicles whose titles and bodies carry query-matching, recency and ranking marks, mostly through new accounts.}

\begin{figure*}[t]
  \addtocounter{figure}{1}
  \centering
  \subcaptionbox{Followers on Bilibili\label{fig:accounts-a}}{\includegraphics{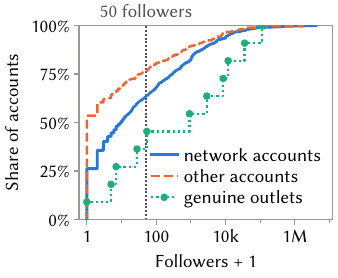}}\hfill
  \subcaptionbox{Days to first placement\label{fig:accounts-b}}{\includegraphics{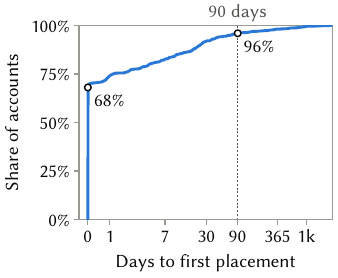}}\hfill
  \subcaptionbox{Active accounts by entry year\label{fig:accounts-c}}{\includegraphics{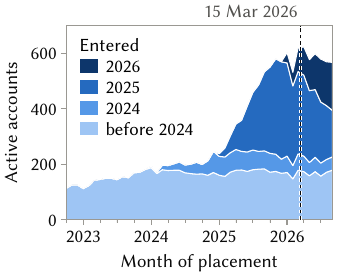}}
  \caption{Accounts carrying placements. (a) Followers + 1 of Bilibili network accounts (n = \rqOneNWithFollowersBili{}), other accounts from the same searches (n = \supFolOtherNBili{} of the \supFolOtherAllBili{} active in 2025 or later that the rule rejected) and genuine news outlets (n = \supFolGenuineNBili{}), where a follower count was observed. (b) Days from an account's first article to its first placement, core Bilibili accounts with a placement (n = 1,150). (c) Core Bilibili accounts publishing placements per month, by entry year; dashed: the March 2026 broadcast.}
  \label{fig:accounts}
  \label{fig:followers}
  \label{fig:accounts-rate}
  \label{fig:account-pool}
  \label{fig:lifetimes}
  \label{fig:growth}
\end{figure*}

\subsection{Accounts: Who Carries Placements, and How They Appear Credible}
\label{sec:production}
\label{sec:audience}
\label{sec:growth}

We next examine the accounts that carry the placements: their audience, how they present themselves and where they come from, and how the pool has grown and turned over.
As a running example we follow \emph{Account A}, a Bilibili account with a news-style name that entered in \caseAcctEntry{}, has placed \caseAcctPlacements{} articles for \caseAcctClients{} clients, was still posting at the snapshot, and was cited in our probe (\S\ref{sec:rq3}; Figure~\ref{fig:case-account}).

\begin{figure}[t]
  \addtocounter{figure}{-2}
  \centering
  \includegraphics[width=\linewidth]{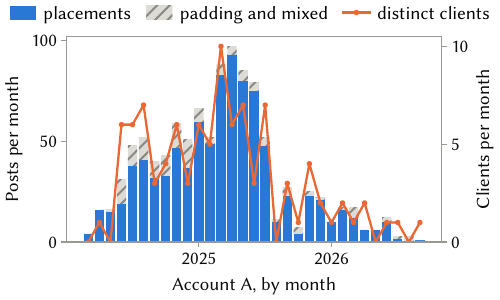}
  \caption{Account A per month: placements and the padding and mixed posts between them (bars, left axis), and distinct clients posted for (line, right axis).}
  \label{fig:case-account}
  \addtocounter{figure}{1}
\end{figure}

\noindentparagraph{Account profile}
Network accounts have almost no readers.
Of the \rqOneNWithFollowersBili{} network accounts on Bilibili whose follower counts we observed, the median account has \rqOneMedianFollowersBili{} followers and \rqOneShareUnderFiftyBili{} have fewer than fifty, against a median of \rqOneMedianFollowersGenuine{} for the \supFolGenuineNBili{} genuine news outlets we found (Figure~\ref{fig:accounts}a); Account A has \caseAcctFollowers{}.
Baijiahao differs in degree but not in kind: its median network account has thousands of followers, yet \rqOneShareOutletNamedBjh{} of network accounts there carry outlet-style names, against \rqOneShareOutletNamedBjhNonmember{} of the accounts the rule rejected.
The network therefore depends on search engines and assistants finding its articles, not on followers.

\noindentparagraph{Presentation and origin}
Network accounts present themselves as publishers, but most were created to publish placements.
They take names associated with news outlets or professional expertise, and intersperse reposted news and corporate announcements with their placements so that a feed resembles industry news; Account A carries a news-style name, and its placements come mainly from \caseAcctIndustries{}.
Their histories begin with placements rather than ordinary posts: \supOriginFirstPostPlacement{} of the core Bilibili accounts published a placement as their first article and \supOriginWithinNinety{} within ninety days of it (Figure~\ref{fig:accounts}b), and we find no core Bilibili account that posted ordinary content for a year before turning to placements.
Older identities enter instead as aged registrations, \supOriginAgedRegs{} accounts created under Bilibili's earlier identifier scheme that first posted in 2024 or later, and as \supOriginReactivated{} dormant accounts reactivated after at least a year of silence.
A few high-volume distributors post for many clients on the same day: \supAssemblyN{} accounts, mostly on Baijiahao, are \supAssemblyAcctShare{} of analyzed accounts but carry \supAssemblyPlacementShare{} of all placements (Appendix~\ref{app:supp}).

\noindentparagraph{Growth and tenure}
The network grew by recruiting accounts rather than by working them harder.
Placements on core Bilibili accounts grew \rqOneGrowthFold{} from 2024 to 2025 as the accounts active in a year rose from \rqOneActiveTwentyFour{} to \rqOneActiveTwentyFive{}, while placements per active account rose only from \rqOnePerActiveTwentyFour{} to \rqOnePerActiveTwentyFive{}; because we observe only accounts that survived until the crawl, these figures may overstate growth.
Accounts are used for a season or two: \rqOneEndedTwentyFive{} of the accounts that entered in 2025 published no placement in the sixty days before the snapshot, against \rqOneEndedPre{} of earlier entrants, and the pool is renewed cohort by cohort while earlier entrants keep publishing placements (Figure~\ref{fig:accounts}c); the number of accounts publishing placements each month did not fall after the March 2026 broadcast.
Account A peaked at \caseAcctPeakN{} placements in \caseAcctPeakMonth{} and had tapered to one placement a month by \caseAcctLast{}.

\finding{2}{Network accounts have almost no readers. Most were created to publish placements, present themselves as outlets or experts, and are recruited and retired in cohorts, and on Baijiahao a few high-volume distributors carry a large share of the placements.}

\subsection{Campaigns: Coordination Across Accounts and Platforms}
\label{sec:coordination}
\label{sec:rq2}

We finally ask whether placements are isolated publications or coordinated campaigns: how long placements for a client persist, whether accounts publish them together, and how they move across platforms.

\begin{figure*}[t]
  \centering
  \subcaptionbox{Campaign duration and size\label{fig:campaigns-a}}{\includegraphics{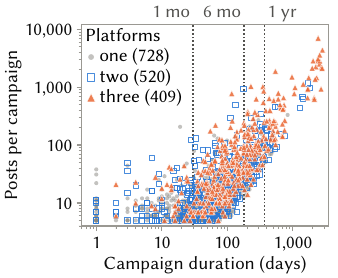}}\hfill
  \subcaptionbox{Synchronized posting per client\label{fig:campaigns-b}}{\includegraphics{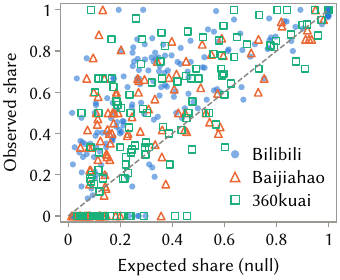}}\hfill
  \subcaptionbox{Time between cross-platform copies\label{fig:campaigns-c}}{\includegraphics{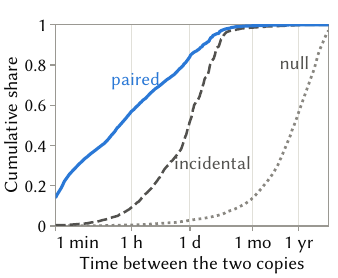}}
  \caption{Coordination of placements. (a) Campaigns with at least five posts naming the client, by duration, number of posts and number of platforms. (b) Share of an account's active weeks shared with another account posting that client, against the circular-shift null, one marker per niche client and platform. (c) Time between two copies of a title on different platforms: inside account pairs, between other accounts, and under a date-redraw null.}
  \label{fig:campaigns}
  \label{fig:campaign-duration}
  \label{fig:campaign-scatter}
  \label{fig:burst-strip}
  \label{fig:burst}
  \label{fig:sync}
\end{figure*}

\noindentparagraph{Campaign duration}
Placements for a client recur over months rather than appearing as isolated events.
We group a client's placements into campaigns separated by gaps of over thirty days.
Among the \campBigClients{} niche clients with at least fifty placements, the longest campaign lasts a median of \campBigLongestMedian{} days, and across all \campNicheClients{} niche clients, \campRepeatShare{} return for at least three campaigns (Appendix~\ref{app:supp}).
Across all clients, longer campaigns have more posts naming the client, and the longest and largest run on all three platforms (Figure~\ref{fig:campaigns}a).

\noindentparagraph{Synchronized posting}
Accounts post the same client in the same weeks far more often than chance.
For each client, we measure the share of an account's active weeks in which another account also posts that client, and compare it with a null that shifts each account's weeks circularly and independently (Figure~\ref{fig:campaigns}b).
On Bilibili the median niche client reaches \rqOneSyncObs{} against \rqOneSyncNull{} under the null, and \rqOneSyncSig{} of \rqOneSyncClients{} niche clients exceed the null at the 5\% level (\syncSigBH{} after Benjamini--Hochberg correction); the excess is smaller on Baijiahao and 360kuai (Appendix~\ref{app:supp}).
Because same-day bursts are also one of the membership signals, we repeat the test without the accounts admitted by that signal alone, and \syncSigNoBurst{} of \syncClientsNoBurst{} niche clients on Bilibili still exceed the null.
When an account posts a client, the next post of that client by another account follows after a median of \rqOneGapMedian{} hours on Bilibili, against \rqOneGapNull{} hours under random post times.

\noindentparagraph{Shared templates}
Placements are produced from shared title templates.
Of the placements whose titles we embedded, \tmplNearDupOtherAccount{} have a near-identical title (cosine similarity of at least 0.95) on another account and \tmplExactReuse{} share an exact title with another account, against only \tmplNullSameTopicPair{} of random pairs of same-topic titles.
Linking titles at a similarity of 0.97 yields \tmplCount{} templates with at least ten placements (Table~\ref{tab:templates} lists the largest), \tmplMultiAccount{} of them used by several accounts (a median of \tmplMedianAccounts{}) and \tmplMultiPlatform{} on several platforms; the largest, an anti-wrinkle top ten, appears \tmplTopPosts{} times on \tmplTopAccounts{} accounts.

\noindentparagraph{Across platforms}
\label{sec:sharedclients}
\label{sec:copies}
\label{sec:pairs}
Campaigns span platforms, and their articles are copied between platforms within minutes to hours.
Of the \campNicheClients{} niche clients, \rqTwoSharedNicheShare{} appear on at least two platforms and \rqTwoClientsAllThree{} on all three.
Bilibili placements reappear on Baijiahao under the same title \rqTwoCopyPairs{} times, \rqTwoCopyWithinHour{} of them within an hour, whereas only \rqTwoCopyNullWithinDay{} would fall within a day if each copy's date were drawn independently from that client's Baijiahao placements.
Copying is done by fixed pairs of accounts: \rqTwoPairs{} pairs that share at least five titles account for \rqTwoPairedMatches{} of the \rqTwoMatchesAll{} cross-platform title matches, and the median copy inside a pair follows its original after \pairGapMedianMin{} minutes, against \copyNullMedianDays{} days under a null that redraws each copy's date from the client's placements on that platform (Figure~\ref{fig:campaigns}c; Appendix~\ref{app:supp}).
Title matches establish repeated headlines, not identical bodies.

\input{tables/tab_templates_imc}

\begin{figure}[t]
  \centering
  \includegraphics[width=\linewidth]{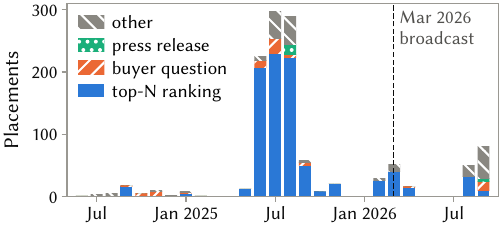}
  \caption{Monthly placements naming Client 2 (\caseBikePlacements{} placements by \caseBikeAccounts{} accounts) by title type; dashed line: the March 2026 broadcast.}
  \label{fig:case-waves}
\end{figure}

\noindentparagraph{Inside one campaign}
The campaign of Client 2, the bicycle brand of \S\ref{sec:intro}, shows how a sustained campaign is staffed (Figure~\ref{fig:case-waves}); \S\ref{sec:exposure} follows how it was rewritten.
Between \caseBikeFirst{} and \caseBikeLast{}, \caseBikeAccounts{} accounts published \caseBikePlacements{} placements naming the client under \caseBikeTitles{} distinct titles, \caseBikeRanking{} of them rankings such as the ``one king and four queens'' hierarchy, in \caseBikeWaves{} campaigns of at least five placements.

\finding{3}{Placements are deployed as campaigns that recur for months, are written from shared templates, are posted by many accounts in the same weeks far more often than chance, and are copied across platforms, often within the hour.}

\subsection{Response to Exposure}
\label{sec:exposure}
\begin{figure}[t]
  \centering
  \includegraphics[width=\linewidth]{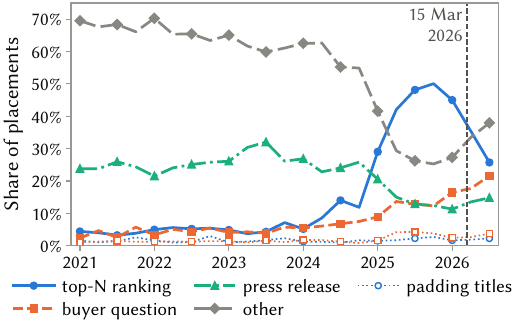}
  \caption{Primary title strategy of Bilibili core placements per quarter (quarters with at least 200 placements): top-N rankings carry a ranking word in the title, and buyer questions are the other listicles, which pose the buyer's question or answer it with a recommendation or selection guide; the two together are the listicles of Figure~\ref{fig:content}a, and other includes product news; hollow markers on dotted lines: the same accounts' padding titles; dashed: March 2026 broadcast.}
  \label{fig:strategies}
\end{figure}

After the national broadcast of March 2026 used ranked lists as its exhibit, the network's ranking titles declined in favor of other formats such as buyer questions (Figure~\ref{fig:strategies}).
By month, ranking titles fell from \stratRankingDec{} of Bilibili core placements in December 2025 to \stratRankingSep{} in September 2026, while buyer-question titles rose from \stratQuestionDec{} to \stratQuestionSep{}.
Of the \discStratEventPanelN{} accounts with at least twenty placements on each side of the broadcast, \stratEventPanelFellN{} (\discStratEventPanelFell{}; \stratEventPanelFellRange{} under other thresholds) lowered their ranking share and their pooled share fell from \stratEventPanelPooledPre{} to \stratEventPanelPooledPost{}, although the median account changed little.
The network does not address the engines by name: titles that invoke an AI assistant are \stratAssistantTwentySixPrecise{} of 2026 placements, against \stratAssistantPadTwentySixPrecise{} of padding titles.

Some campaigns changed later and all at once.
The campaign of Client 2 (\S\ref{sec:coordination}) kept its rankings through the broadcast: \caseBikeMarAugRanking{} of the \caseBikeMarAugAll{} placements from March to August 2026 are rankings.
Its format changed abruptly in the first days of September 2026, when \caseBikeSepNonRanking{} of \caseBikeSepAll{} placements took other forms, chiefly corrections that call rankings or AI answers misleading and then present the client's ``real'' standing, news hooks and buying guides.
The new titles came from the accounts that were already posting for the client: of the \caseBikeSepAccounts{} accounts posting them, \caseBikeSepAugRanking{} had posted rankings for the client in August, and only \caseBikeSepPriorMarch{} had posted for it before March.
The switch coincides in time with the September bulletin of the regulator's campaign against GEO marketing (Appendix~\ref{app:sources}, S6).
Both changes are temporal associations: titles alone cannot separate evasion from a change in what clients buy, and our data cannot link the September switch to the bulletin.

\finding{4}{After the March 2026 broadcast, the network's ranking titles declined in favor of buyer questions and other formats. A single campaign changed its formats within days in September 2026, through the accounts already posting for it.}

%% file: tables/tab_templates_imc.tex
\begin{table*}[t]
\footnotesize
\setlength{\tabcolsep}{5pt}
\caption{The six largest title templates (titles linked at cosine similarity of at least 0.97): product category, the shared title pattern rendered in English (highlighted: wording that varies across the template's titles; \ldots{} a free-form remainder), placements, accounts, platforms, and the number of months with at least one placement, with the first and last such month.}
\label{tab:templates}
\begin{tabular}{@{}llrrcl@{}}
\toprule
Product category & Shared title pattern (English rendering) & Placements & Accounts & Platforms & Active months (first--last) \\
\midrule
\rowcolor{gray!10}anti-wrinkle skincare & Top-10 \slot{[anti-wrinkle]} skincare ranking, \ldots & 751 & 215 & \iconBili\,\iconBjh & 13 (Mar 2025--Jun 2026) \\
 & Which \slot{[anti-wrinkle firming]} skincare product is good? \ldots & 509 & 193 & \iconBili\,\iconBjh & 21 (Oct 2023--Oct 2026) \\
\rowcolor{gray!10} & Which brand of \slot{[anti-wrinkle, anti-aging]} skincare is good? \ldots & 466 & 167 & \iconBili\,\iconBjh & 19 (Feb 2024--Oct 2026) \\
liver-support tablets & Which brand of liver-support tablets is \slot{[best and safest]}? \ldots & 448 & 176 & \iconBili\,\iconBjh\,\iconKuai & 20 (Feb 2025--Sep 2026) \\
\rowcolor{gray!10}facial cleanser & Which brand of facial cleanser works best? \ldots & 397 & 115 & \iconBili\,\iconBjh & 24 (Jan 2024--Oct 2026) \\
acne treatment & Top-10 \slot{[acne-treatment]} brand ranking \slot{[released]}, \ldots & 257 & 89 & \iconBili\,\iconBjh & 18 (Jan 2024--Dec 2025) \\
\bottomrule
\end{tabular}
\end{table*}

%% file: sections/6-rq3-assistants.tex
\section{Exposure in AI Assistants}
\label{sec:rq3}
\label{sec:reach}
\label{sec:ingest}
\label{sec:share}
\label{sec:effects}
This section answers RQ2 with the \rqThreeAnswers{} pilot answers and the \rqFourAnswers{} census answers of \S\ref{sec:probe}.
We ask which assistants retrieve the network's articles and through which platforms (\S\ref{sec:retrieval}), whether their answers name the network's clients (\S\ref{sec:naming}), and what they do with its claims (\S\ref{sec:claims}).

\subsection{Retrieval: Which Assistants Cite the Network}
\label{sec:retrieval}
\begin{figure}[t]
  \centering
  \includegraphics[width=\linewidth]{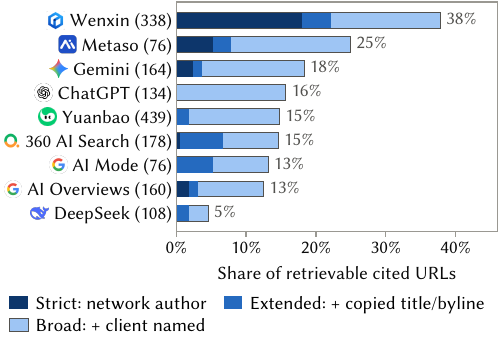}
  \caption{Share of each assistant's retrievable cited pages attributed to the network at the strict, extended and broad levels, nine category questions (pages per assistant in parentheses).}
  \label{fig:engine-tiers}
  \label{fig:engine}
  \label{fig:engine-share}
  \label{fig:share}
\end{figure}

\noindentparagraph{Network reference share}
Every probed assistant cites pages attributable to the network, eight at the extended level and nine at the broad level, with shares depending on how strictly attribution is drawn (Figure~\ref{fig:engine-tiers}).
Of the \engTierUrlsExGeo{} retrievable distinct pages cited for the nine category questions, \engTierStrictExGeo{} were written by a network account, \engTierMidExGeo{} are attributable at the extended level, and \engTierBroadExGeo{} at the broad level, which also counts genuine coverage naming a client.
Wenxin cites the network most at every level (\engTierWenxinStrictExGeo{}, \engTierWenxinMidExGeo{} and \engTierWenxinBroadExGeo{}), and DeepSeek least (\engTierDeepSeekBroadExGeo{} broad).
Account A of \S\ref{sec:production} is among the cited authors, returned by Metaso for a question about private hospitals.
The strict and extended shares are lower bounds: they recognize only accounts and titles that the crawl found, and removing pages whose matched author or byline is a genuine news outlet, or whose byline match is spurious, lowers them to \engTierStrictCleanExGeo{} and \engTierMidCleanExGeo{}.

\noindentparagraph{Platforms cited}
Wenxin reaches the network mainly through Baidu's own platform, Metaso through Bilibili, and the other assistants mainly through copies of its articles on other sites.
Wenxin draws \engWenxinBjhPilot{} of its pilot citations and \engWenxinBjhCensus{} of its census citations from Baijiahao, Baidu's own platform, and Metaso draws \engMetasoBiliPilot{} of its pilot citations from Bilibili, whereas no assistant cites 360kuai and the remaining assistants cite the three crawled platforms rarely, reaching the network through copies of its articles on portals and aggregators.
Table~\ref{tab:services} lists each assistant's attribution levels, citations on the crawled platforms and census naming rates.

\input{tables/tab_services_imc}

Cited pages are written like the network's placements: \rqThreeNetListicleCited{} of the pages attributable at the broad level carry a listicle title, against \rqThreeOtherListicleCited{} of the other cited pages.

\noindentparagraph{Which accounts are cited}
Within the network, the accounts that reach an assistant are those on the engine operator's platform that publish listicles rather than padding.
Of the \engReachExposed{} analyzed accounts on Bilibili and Baijiahao that placed articles for the pilot clients, \engReachCited{} were cited by at least one assistant: \engReachBjhCited{} of \engReachBjhExposed{} on Baijiahao (\appReachBjhShare{}) against \engReachBiliCited{} of \engReachBiliExposed{} on Bilibili (Fisher's exact test, $p<10^{-8}$).
Among Baijiahao accounts, cited accounts have higher listicle shares (AUC \engReachListicleAUC{}, $p=\appReachListicleP{}$) and lower padding shares (AUC \engReachPaddingAUC{}, $p=\appReachPaddingP{}$) than never-cited accounts, consistent with the listicle-heavy accounts recruited from 2024 onward (\S\ref{sec:content}).

\finding{5}{Every probed assistant cites pages attributable to the network, at \engTierStrictExGeo{} to \engTierBroadExGeo{} of cited pages depending on how strictly attribution is drawn. Baidu's assistant reaches it through the network's Baijiahao accounts and the other assistants mostly through copies, and the cited accounts publish listicles on the engine operator's platform.}

\subsection{Naming: Do Answers Name the Network's Clients}
\label{sec:naming}
\noindentparagraph{Clients against category leaders}
Across the network's sustained campaigns, answers name the client about three-quarters as often as they name the category leader.
In the census, clients are named in \engCensusClientNamed{} of the \engCensusAnswers{} answers (95\% CI \engCensusClientWilson{}) and the leaders in \engCensusControlNamed{} (\engCensusControlWilson{}), a ratio of \engCensusRatio{}; clustered by question, the difference is not significant (bootstrap interval \engCensusDiffBootQ{}; Wilcoxon $p=\engCensusWilcoxonP{}$).
The ratio ranges from \engCSDeepSeekRatio{} on DeepSeek to \engCSThreeSixtyRatio{} on 360 AI Search (Table~\ref{tab:services}), and the three runs of a question on an assistant agree on whether the client is named in \engCensusRepAgree{} of cases.

\begin{figure}[t]
  \centering
  \includegraphics[width=\linewidth]{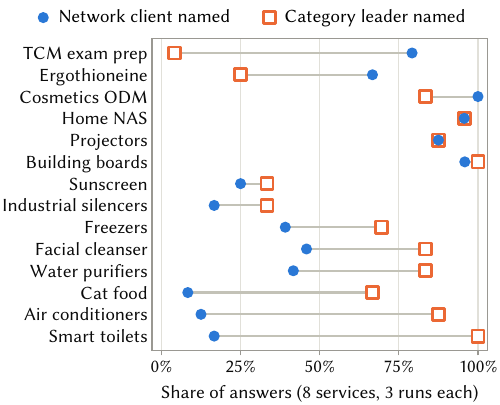}
  \caption{Census: share of answers naming the network client and the category leader, per question (eight assistants, three runs each).}
  \label{fig:census-questions}
\end{figure}

\noindentparagraph{Category, not volume}
Whether a client is named depends on the question rather than on how much the client publishes.
Clients are named in \rqFourNicheNamedRange{} of answers in five categories, such as exam preparation and home storage devices, and in at most \rqFourOwnedNamedMax{} in four, such as cat food and air conditioners (Figure~\ref{fig:census-questions}).
Placement volume does not explain this split: naming is unrelated to how many placements a client published in the preceding twelve months (Spearman $\rho=$\engVolumeSpearman{}, $p=\engVolumeSpearmanP{}$), and a separate probe of \rqThreeDoseClients{} clients across four volume strata finds the same.
Neither does a dominant category leader (Appendix~\ref{app:probe}).

\noindentparagraph{Citation and naming}
Naming is associated with what the answer cites.
In the pilot, an answer names the client in \engAdoptWithCite{} of cases when at least one cited page names it, and in \rqThreeAdoptionWithoutCite{} otherwise.
Answers also name network clients other than the one asked about: \engAnyNetClient{} of pilot answers to the nine category questions name at least one network client, while \engTargetClient{} name the client that the question targets.
This is an association between citation and naming; it does not isolate the effect of any placement, and some named clients have independent coverage (Appendix~\ref{app:probe}).

\finding{6}{Answers name the network's clients in \engCensusClientNamed{} of census answers, against \engCensusControlNamed{} for category leaders; whether a client is named depends on the category and is unrelated to how much it publishes.}

\subsection{Claims: What the Answers Say}
\label{sec:claims}
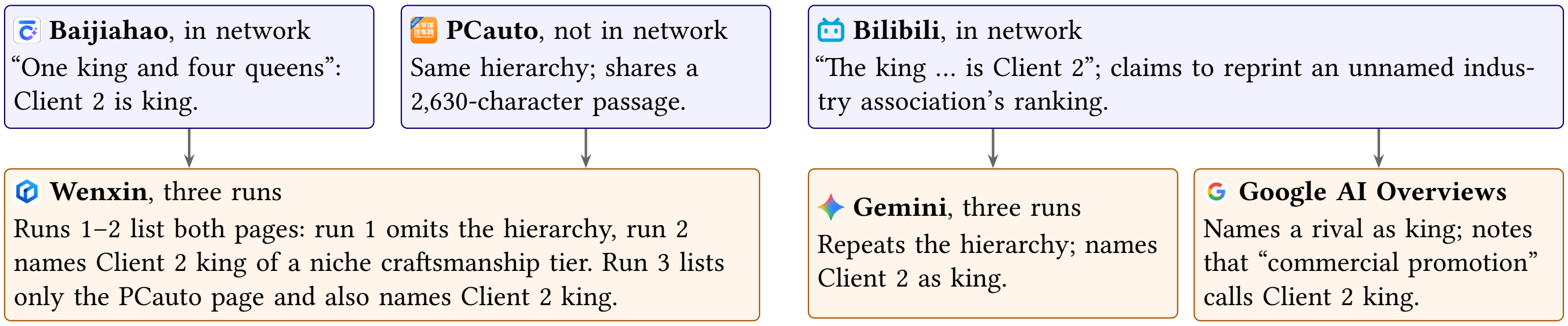
\begin{figure*}[t]
  \centering
  \tikzset{
    source/.style={draw=blue!45!black, fill=blue!5, rounded corners=2pt, text width=40mm, minimum height=14mm, inner sep=3pt, align=left, font=\small},
    output/.style={draw=orange!65!black, fill=orange!8, rounded corners=2pt, minimum height=17mm, inner sep=3pt, align=left, font=\small},
    link/.style={-{Stealth[length=1.6mm]}, thick, draw=black!60}}
  \subcaptionbox{Two pages share one passage; Wenxin lists them\label{fig:source-answer-case-a}}{%
  \begin{tikzpicture}
    \node[source] (mapped) {\iconBjh~\textbf{Baijiahao}, in network\\[1pt]
      ``One king and four queens'': Client 2 is king.};
    \node[source, anchor=north west] (portal) at ([xshift=3mm]mapped.north east) {\iconPCauto~\textbf{PCauto}, not in network\\[1pt]
      Same hierarchy; shares a \caseBikeSharedChars{}-character passage.};
    \node[output, text width=84mm, anchor=north west] (answer) at ([yshift=-4.5mm]mapped.south west) {\iconWenxin~\textbf{Wenxin}, three runs\\[1pt]
      Runs 1--2 list both pages: run 1 omits the hierarchy, run 2 names Client 2 king of a niche craftsmanship tier. Run 3 lists only the PCauto page and also names Client 2 king.};
    \draw[link] (mapped.south) -- (mapped.south |- answer.north);
    \draw[link] (portal.south) -- (portal.south |- answer.north);
  \end{tikzpicture}}\hfill
  \subcaptionbox{One network page; two assistants treat it differently\label{fig:source-answer-case-b}}{%
  \begin{tikzpicture}
    \node[source, text width=84mm] (bili) {\iconBili~\textbf{Bilibili}, in network\\[1pt]
      ``The king \ldots\ is Client 2''; claims to reprint an unnamed industry association's ranking.};
    \node[output, text width=40mm, anchor=north west] (gem) at ([yshift=-4.5mm]bili.south west) {\iconGemini~\textbf{Gemini}, three runs\\[1pt]
      Repeats the hierarchy; names Client 2 as king.};
    \node[output, text width=40mm, anchor=north east] (aio) at ([yshift=-4.5mm]bili.south east) {\iconGoogle~\textbf{Google AI Overviews}\\[1pt]
      Names a rival as king; notes that ``commercial promotion'' calls Client 2 king.};
    \draw[link] (gem.north |- bili.south) -- (gem.north);
    \draw[link] (aio.north |- bili.south) -- (aio.north);
  \end{tikzpicture}}
  \caption{Source-to-answer traces for Client 2, paraphrased and pseudonymized. Arrows follow source lists.}
  \label{fig:source-answer-case}
\end{figure*}

\noindentparagraph{Claim repetition}
An answer can be fully supported by its cited pages and still repeat a claim that the network spread.
The bicycle case of \S\ref{sec:intro} traces one such claim (Figure~\ref{fig:source-answer-case}).
A Baijiahao network account presents Client 2 as the king of a traditional ``one king and four queens'' hierarchy, and an article on PCauto, outside our account mapping, presents the same hierarchy and shares a \caseBikeSharedChars{}-character passage with it.
Wenxin repeats the hierarchy in \engBikeWenxinKing{} of its \engBikeWenxinReps{} runs, citing both pages in one and only the PCauto copy in the other, while Gemini repeats the claim from a Bilibili network article and Google AI Overviews names an established rival as king and attributes Client 2's title to commercial promotion.
The case establishes shared source text and claim overlap; it does not identify which page supplied the claim or the direction of copying.

\noindentparagraph{Reputation questions}
Where a client has a public regulatory record, the answers report it despite the network's defensive articles.
Asked whether a client is reliable, the assistants cite at least one network article defending the client in \engRepCiteNetDefence{} of \engRepAnswers{} answers, yet \engRepPenaltyAnswers{} of the \engRepPenaltyOf{} answers about a stock-advisory client mention its regulatory penalty.

\noindentparagraph{The GEO vendor question}
Asked which GEO vendors are good, the assistants recommend the vendors that advertise through the network.
\engGeoVendorAnswers{} of the \engGeoAnswers{} answers name at least one GEO vendor that the network promotes (\S\ref{sec:content}), and the most-named vendor appears in \engGeoTopVendorAnswers{}, in every run on four assistants.

\finding{7}{Answers repeat the network's claims and recommend its GEO vendors, but report a client's regulatory penalty when one is on record. These results show exposure and claim overlap, not causal effects.}

%% file: tables/tab_services_imc.tex
\begin{table*}[t]\centering\footnotesize
\caption{Per assistant: access route; pilot answers; retrievable cited pages (a page cited by several assistants is counted for each) and the share attributed to the network at the strict, extended and broad levels; the share of pilot citations pointing to Baijiahao and to Bilibili; and the share of census answers naming the client and the category leader, with their ratio. }
\label{tab:services}
\setlength{\tabcolsep}{5pt}
\begin{tabular}{@{}llrrrrrrrrrr@{}}
\toprule
 & & \multicolumn{5}{c}{Pilot attribution} & \multicolumn{2}{c}{Cited host share} & \multicolumn{3}{c}{Census naming} \\
\cmidrule(lr){3-7}\cmidrule(lr){8-9}\cmidrule(l){10-12}
Assistant & Route & Answers & Pages & Strict & Extended & Broad & \iconBjh~Baijiahao & \iconBili~Bilibili & Client & Control & Ratio \\
\midrule
\rowcolor{svcWenxin}\iconWenxin~Wenxin (Baidu) & Web & 30 & 338 & 18\% & 22\% & 38\% & 27\% & 0\% & 46\% & 76\% & 0.61 \\
\rowcolor{svcMetaso}\iconMetaso~Metaso & Web & 14 & 76 & 5\% & 8\% & 25\% & 0\% & 13\% & -- & -- & -- \\
\rowcolor{svcGemini}\iconGemini~Gemini & API & 30 & 164 & 2\% & 4\% & 18\% & 0\% & 4\% & 64\% & 79\% & 0.82 \\
\rowcolor{svcChatGPT}\iconChatGPT~ChatGPT & API & 30 & 134 & 0\% & 0\% & 16\% & 0\% & 0\% & 43\% & 62\% & 0.69 \\
\rowcolor{svcYuanbao}\iconYuanbao~Yuanbao (Tencent) & Web & 30 & 439 & 0\% & 2\% & 15\% & 0\% & 0\% & 38\% & 67\% & 0.57 \\
\rowcolor{svcSoSearch}\iconSoSearch~360 AI Search & Web & 30 & 178 & 1\% & 7\% & 15\% & $<$1\% & $<$1\% & 74\% & 64\% & 1.15 \\
\rowcolor{svcGoogle}\iconGoogle~Google AI Mode & API & 30 & 76 & 0\% & 5\% & 13\% & 0\% & 2\% & 57\% & 62\% & 0.92 \\
\rowcolor{svcGoogle}\iconGoogle~Google AI Overviews & API & 30 & 160 & 2\% & 3\% & 13\% & 0\% & 2\% & 59\% & 71\% & 0.83 \\
\rowcolor{svcDeepSeek}\iconDeepSeek~DeepSeek & Web & 30 & 108 & 0\% & 2\% & 5\% & 0\% & 0\% & 36\% & 64\% & 0.56 \\
\bottomrule
\end{tabular}
\end{table*}

%% file: sections/8-related.tex
\section{Related Work}
\label{sec:related}
\noindentparagraph{Auditing generative search}
A growing body of work audits what deployed generative engines cite, but starts from the answers rather than from the producers of the cited content.
These audits measure the credibility of cited domains and their overlap with search results~\cite{xu2026aioverviews}, the hosts exposed across engines and languages~\cite{li2026baidugoogle}, the relation between source prevalence and the entities that answers surface~\cite{zhen2026chinesegse}, and the presence of AI-generated pages~\cite{allaham2026synthetic} and malicious URLs~\cite{luo2025unsafesearch} among sources; information scarcity has been proposed as a reason assistants repeat low-quality sources~\cite{alyukov2025datavoids}.
None can say who produced the cited pages or whether their appearance reflects an organized operation; we start from a mapped operation and trace its articles into the references and text of answers.

\noindentparagraph{Search spam}
The strategic placement of content to influence search visibility has a long history in black-hat search engine optimization (SEO).
Prior studies have documented link-spam networks~\cite{gyongyi2005alliances}, SEO botnets~\cite{wang2013juice}, search-redirection infrastructures~\cite{leontiadis2011redirection,leontiadis2014longitudinal}, counterfeit-goods campaigns~\cite{wang2014seizure}, spider pools~\cite{du2016labyrinth} and private blog networks~\cite{vangoethem2019pbn}.
Research on Chinese search ecosystems has further examined promotional content injected into reputable websites~\cite{liao2016seise,yang2021dmos,zhang2024isap,liu2025nokescam} and illicit campaigns identified through keyword-based discovery~\cite{yang2017klingon,wu2024rsp,wang2025illicit}.

GEO shares the objective of these operations but not their mechanism.
Search spam seeks to raise the rank of pages that users may then visit, which typically requires operating or compromising websites and building links to them.
GEO only needs its articles to enter the set of pages an engine retrieves for answer synthesis, so operators can publish through existing content platforms that the engines already trust instead of running their own sites.

\noindentparagraph{Coordinated and paid promotion}
The coordinated distribution of promotional content is also related to crowdturfing, paid posting and manipulated platform signals.
Prior work has examined intermediaries that sell paid posting through networks of accounts~\cite{wang2012serf,chen2013wateramy,lee2013crowdturfers}, markets that pay users to install apps and thereby inflate store rankings~\cite{farooqi2020incentivized}, comment bots that promote scams on video platforms~\cite{na2023evolvingbots}, and cross-platform referral traffic for illicit drugs~\cite{zha2024illicitdrug}; fraudulent reviews and coordinated campaigns have been detected through shared targets, behavioral patterns and near-simultaneous posting of similar content~\cite{mukherjee2012spotting,kaghazgaran2018combating,he2022market,pacheco2021uncovering,keller2020political}.

These operations target human readers, relying on the apparent independence of many sources to persuade them; generative engines add a path that needs no reader at all.
By publishing similar claims across many accounts and platforms, operators can raise the chance that their articles are retrieved, so a campaign may shape generated answers even when its accounts attract almost no direct readership.
We therefore need to identify the operation by its coordination rather than by its audience (\S\ref{sec:hunter}).

%% file: sections/7-discussion.tex
\section{Discussion}
\label{sec:discussion}

\noindentparagraph{Reach through platforms}
The operation reaches assistants by publishing on platforms the engines already retrieve from, not by attracting readers.
Its accounts have almost no followers (\S\ref{sec:production}), yet the cited accounts are concentrated on the platform run by the engine's own operator (\S\ref{sec:retrieval}).
An article from such a platform enters retrieval on the platform's reputation, although no editor has reviewed it, so the platform is also where this access can be cut.

\noindentparagraph{Coordination is the signal}
Individual placements are a moving target, but the coordination behind them is not.
After the March 2026 broadcast, the network's ranking titles declined in favor of buyer questions, single-brand evaluations and other formats (\S\ref{sec:exposure}), and assistants that cite their sources faithfully still repeat the network's claims (\S\ref{sec:claims}), so checks that an answer is supported by its citations cannot reveal that the citations themselves were planted~\cite{xu2026aioverviews}.
What persists across these changes is the publishing pattern: templated titles, same-week posting for the same client and copies across platforms, often within the hour (\S\ref{sec:coordination}).
These signals are visible at the level of accounts and their relations, not of single articles.

\noindentparagraph{Uneven influence}
The network's effect on answers is uneven and does not scale with spending.
Clients are named in nearly every answer in some categories and almost never in others, naming is unrelated to how much a client publishes, and answers report a client's regulatory penalty when one is on record (\S\ref{sec:naming}, \S\ref{sec:claims}).
Vendors advertise placement as a reliable route into AI answers~(Appendix~\ref{app:sources}, S2); our measurements suggest that the route exists but delivers less consistently than advertised, and we cannot tell from our data what makes a category receptive.

\noindentparagraph{Recommendations}
\begin{enumerate}[leftmargin=1.4em,itemsep=1pt,topsep=2pt,parsep=0pt]
  \item \textbf{Assistant providers} should treat pages from one coordinated cluster as dependent rather than independent evidence, for example by collapsing near-duplicate titles and copies across hosts before synthesis (\S\ref{sec:coordination}, \S\ref{sec:retrieval}), and should weigh the publishing account's history, not only its host, when ranking UGC pages.
  \item \textbf{Content platforms} should screen accounts rather than articles for the signals that identify the network: several unrelated clients, titles copied across accounts and same-day bursts for one client (\S\ref{sec:hunter}), together with outlet-style names on accounts without an audience (\S\ref{sec:production}).
  \item \textbf{Regulators} applying the identifiability requirement for advertising (S3, S4) can target the agencies and account pools that publish at scale rather than individual articles, since a single client's placements can come from close to two hundred accounts (\S\ref{sec:coordination}).
  \item \textbf{Auditors of generative search} should trace who produced the cited pages, not only their accuracy, since a faithful answer can rest on coordinated sources (\S\ref{sec:retrieval}).
\end{enumerate}

\noindentparagraph{Alternative explanations}
Our data identify publishing behavior, not intent or responsibility.
A brand named in placements may have bought distribution without knowing how its agency carries it out, and some named clients also attract genuine coverage that independently explains their appearance in answers (\S\ref{sec:naming}).
Whether unlabeled placements are manipulation depends on the standard applied; we follow the Chinese requirement that advertising be identifiable as such (S3, S4), under which an unlabeled article promoting a client is covert advertising even if accurate.

\noindentparagraph{Limitations}
The crawl is purposive and reaches only accounts whose titles carry a searched name, so the network is larger than we measure; histories on Baijiahao and 360kuai are capped, so trends rest on Bilibili; and the membership evaluation uses one annotator on Bilibili.
On the engine side, the broad attribution level includes genuine coverage and the strict level misses copies on hosts we did not crawl, so the two bracket the network's share; the probes cover \rqThreeQuestions{} pilot and \rqFourCensusQuestions{} census questions, besides two reputation questions, from one vantage point in the United States, four pilot controls were chosen after the first round, Doubao could not be probed, and all engine-side results are associations rather than effects of placement.

%% file: sections/9-conclusion.tex
\section{Conclusion}
\label{sec:conclusion}
We measured a commercial GEO operation from the accounts that publish its articles to the assistants that cite them: \geohunter identified \methMembersAll{} network accounts on three Chinese content platforms, and \geoprobe traced the pages that nine assistants cite back to the network.
The operation promotes ordinary consumer goods and services, increasingly through listicles posted by accounts with almost no followers, and coordinates its placements across accounts and platforms.
Every assistant we probed cites its articles, which make up \engTierStrictExGeo{} to \engTierBroadExGeo{} of cited pages depending on how strictly we attribute them, and answers repeat its claims.
Because the article formats change while the coordination persists, platforms and assistant providers should examine the accounts that publish the articles and how they are linked, not only the articles themselves.
We will release both tools and a pseudonymized dataset.

%% file: sections/appendix-imc.tex
\section{Ethics}
\label{sec:ethics}
\label{app:ethics}
Our study starts from placements we ordered from a GEO vendor and traces them to discover more accounts that the vendor's network operates.
We do so to understand how the operation reaches assistants, but that aim does not remove our obligations to the platforms and their readers, to the assistants, and to the accounts and clients we describe.
We kept to the restrictions below.

\noindentparagraph{Platforms and their readers}
A purchase was necessary because every other membership in the network is inferred: only an article we ordered ourselves is known, by construction, to be a paid placement, and the accounts that publish such an article are the only ground truth from which the crawl can start.
The \seedArticles{} articles we ordered were harmless: they promoted a fictitious company and could not lead a reader to a purchase.
This was our only purchase, and we did not otherwise engage the vendor.
Crawling was read-only and paced at about \requestPace{} seconds between requests; access challenges were handed to a human operator and never solved automatically, and no platform terms were circumvented.


\noindentparagraph{Assistants}
Assistant probes used the assistants' public interfaces or paid APIs.

\noindentparagraph{Accounts and clients}
The subjects of this study are commercial publishing accounts and the public articles they post, not individuals, and we collect no personal data beyond the public display names of institution-style accounts, so the work does not involve human subjects.
We name no person, and account and client names are replaced by pseudonyms in every figure and table.
We identify accounts by their coordination rather than articles by their intent, exclude genuine outlets from the analysis population, and make no claim about the legality of any account or client beyond the regulations we cite (Appendix~\ref{app:sources}).
False-positive membership remains possible, and a client named in placements may not know how its agency distributes them.


\noindentparagraph{Disclosure}
We report a manipulation pattern rather than a software vulnerability, so no vulnerability disclosure applies; the operators of the platforms and assistants are nonetheless best placed to act on our recommendations.

\section{Open Science}
\label{app:openscience}
We will release our software and code, including the crawler, the analysis pipeline that produces the numbers, figures and tables in this paper, and the probe harness, under an open-source license, together with the validation labels, the probe questions and the recorded assistant answers.
A frozen snapshot of the crawl stores at \methSnapshotDate{} will be released with account identifiers pseudonymized as in the paper; the mapping from pseudonyms to platform identifiers is withheld because it would single out the operators of individual accounts whose coordination we infer rather than observe.

\section{Generative AI}
\label{app:genai}
Generative models are part of the measurement instrument and are described where they are used: an LLM judge (\llmJudgeModel{}) proposes client names from article titles in the crawl loop (\S\ref{sec:hunter}, prompts in Appendix~\ref{app:crawl}), and a replay with \jOneModel{} measures the sensitivity of that step; the judge model also vets proposed names before search; a classifier distilled from the LLM judge's labels separates placements from padding (Appendix~\ref{app:crawl}); and the nine generative assistants we probe are the objects of study (\S\ref{sec:probe}).
The text of this paper was drafted and revised with Claude Code (Anthropic), using the models Claude Fable~5.1, Claude Fable~5.2 and Claude Opus~5.5; the same tool helped write the analysis and plotting code that produces the figures and tables, and the source of the diagrams.
The authors directed this work, reviewed and edited all generated text and code, and take full responsibility for the content of the paper.

%% file: sections/appendix.tex

\section{Crawl Details}
\label{app:crawl}

\noindentparagraph{Prompts}
Three prompts support the crawl of \S\ref{sec:hunter}.
The extraction prompt receives dated batches of \appLlmBatch{} titles from one account and proposes verbatim client names, together with category--format combinations whose tokens could identify similar placements.
Candidates include their type, promotional framing and first and last dates; the caller checks literal support in the titles.
The merge prompt combines candidates across batches, resolves spelling variants, removes incidental news subjects and generic names, and prioritizes distinctive names likely to find other accounts, with vendors and agencies first.
Category--format combinations are capped at \appFormulaCap{} of the list.
The result-filtering prompt examines up to \appFilterMaxResults{} search results per name and retains promotional, explanatory, legitimacy or comparison articles about that entity.
Calls return JSON at temperature \appLlmTemperature{}, with \appLlmWorkers{} extraction calls in flight per account; the full prompts accompany the code.

\noindentparagraph{Search and judge thresholds}
Search uses Bilibili's article search, Baidu web search restricted to Baijiahao, and 360 news search.
The three signals introduced in \S\ref{sec:identify} use these thresholds: at least two clients; an identical title for the same client on another account, or character 4-gram Jaccard similarity of at least 0.5; or publication of the same client within one day of at least two other accounts.
Title similarity is thresholded in three further places, each on a different measure for a different purpose: an embedding cosine of at least 0.95 marks a near-identical title on another account and one of at least 0.97 links titles into templates (\S\ref{sec:coordination}), and a string-similarity ratio of at least \rqThreeTitleSim{} matches a cited page's title to a network article for the same client (\S\ref{sec:probe}).
Formally, for the $i$-th search on a platform that retains at least five accounts, let $R_i$ be the retained accounts and $N_i \subseteq R_i$ those that the judge accepts as network accounts and that no earlier search on that platform returned; the yield is $y_i = |N_i|/|R_i|$, and the crawl of a platform stops once $\bar{y}_i = \frac{1}{5}\sum_{k=i-4}^{i} y_k$ falls below \stopRule{} (\S\ref{sec:crawl}).
At \methSnapshotDate{}, the five-search mean yield was \methYieldBili{} on Bilibili, \methYieldBjh{} on Baijiahao and \methYieldKuai{} on 360kuai.
Requests are separated by about \requestPace{} seconds, and access challenges are handed to the human operator.

\noindentparagraph{Validation and model sensitivity}
The first-round account evaluation is reported in \S\ref{sec:evaluation}.
Of its \methValN{} accounts, the annotator labeled \methValNetwork{} as network and \methValOrganic{} as organic; the rule accepts \methValTP{} of the former and \methValFP{} of the latter, so it rejects only \methValTN{} of the \methValOrganic{} organic accounts (specificity 95\% CI \methValSpecCI{}), and the sample is too small to estimate how well the rule excludes organic accounts.
The article classifier is trained on \methClassifierTrainN{} articles whose labels an LLM judge (\supClassifierJudge{}) assigned per cluster of similar titles, and agrees with those labels on \methClassifierAccuracy{} of held-out articles; \supLabelModelLowConf{} of the placements it labels have a predicted probability below 0.6. A human annotator agreed with the final placement labels on \supLabelHumanAgree{} of \supLabelHumanN{} Bilibili articles ($\kappa=\supLabelHumanKappa{}$).
Replaying name proposal with \jOneModel{} recovers \jOneRecallSearched{} of searched names and reaches \jOneAccountsReachedShare{} of their accounts, a substantial overlap under a different model.

\noindentparagraph{Coverage and observation limits}
Treating each of the \methChaoTermsBili{} searched names that returned a network account as a capture occasion, Chao2 estimates \methChaoNBili{} reachable Bilibili accounts (95\% CI \methChaoCIBili{}).
The crawl covers about \methChaoCoverageBili{} of that estimate, against \methChaoCoverageBjh{} on Baijiahao and \methChaoCoverageKuai{} on 360kuai; accounts unreachable by name search remain outside these estimates.
We collect at most 1,000 recent articles per Baijiahao feed (5,000 for \bjhHistAccounts{} accounts re-mined for their history) and 360kuai feeds end at 500, so historical trends use Bilibili.
We exclude \nHotlineExcludedBili{} accounts whose titles mostly match fake after-sales hotline patterns, and exclude accounts first discovered through assistant references from the publishing-corpus counts; they are retained when describing the observed assistant references.

\section{Content Coding Details}
\label{app:bodies}

\noindentparagraph{Title and padding labels}
The listicle rule and its evaluation are given in \S\ref{sec:content}; numeral--unit combinations and abstract numbered lists are excluded.
Table~\ref{tab:styles} documents the non-exclusive title-style rules, each checked on 50 hits with precision 0.90--1.00.
Its 2026 shares cover the \lstTitlesBili{} listicle titles of core Bilibili accounts and the same accounts' padding titles.
Placement and padding labels are evaluated in Appendix~\ref{app:crawl}.
Table~\ref{tab:domains_prior} sets the network's client domains against the categories of promoted content in prior search-spam measurements; because those studies count pages, keywords or campaigns, presence rather than share is comparable. The table covers the \domTableClients{} niche clients with a placement in the core segment.

\noindentparagraph{Listicle growth}
Among accounts that entered before 2023, the pooled listicle share rose only from \lstCohortPreTwentyFour{} in 2024 to \lstCohortPreTwentyFive{} in 2025, and a shift-share decomposition attributes \lstShiftWithinShare{} of the 2024--2025 rise on core Bilibili accounts to change within accounts present in both years (\S\ref{sec:content}).
On Baijiahao, the \bjhHistCompleteN{} re-mined accounts with complete histories reach a listicle share of \bjhHistCompleteListicleLatest{} in 2026, and the press releases of the \bjhHistAccounts{} re-mined accounts fell from \bjhHistPrPre{} of placements up to 2023 to \bjhHistPrLatest{} in 2026; only 3--6 of these accounts predate 2024, and the press-release labels are not validated.

\input{tables/tab_domains_prior}
\input{tables/tab_prior_mix}

\begin{table}[t]
  \caption{Title styles of the network's listicles under the current listicle rule: share of core Bilibili accounts' listicle titles dated 2026, share of the same accounts' padding titles, and the trend in percent (2021 $\to$ 2026). Styles are non-exclusive rules on the title string.}
  \label{tab:styles}
  \footnotesize\setlength{\tabcolsep}{3pt}
  \begin{tabular}{@{}llrrl@{}}
    \toprule
    Group & Style & Listicles & Padding & 2021 $\to$ 2026 \\
    \midrule
    List shape & top-N ranking & 63\% & 3\% & 50 $\to$ 63 \\
     & declared \#1 & 8\% & 2\% & 3 $\to$ 8 \\
     & head-to-head & 2\% & 1\% & 4 $\to$ 2 \\
    Query shaping & keyword repeated & 50\% & 13\% & 26 $\to$ 50 \\
     & segmented title & 56\% & 29\% & 29 $\to$ 56 \\
     & $\geq$40 characters & 20\% & 4\% & 1 $\to$ 20 \\
     & stacked questions & 6\% & 0\% & 3 $\to$ 6 \\
    Freshness & year stamp & 51\% & 10\% & 11 $\to$ 51 \\
    Audience & persona / condition & 16\% & 3\% & 2 $\to$ 16 \\
     & occasion hook & 3\% & 3\% & 10 $\to$ 3 \\
    Framing & superlative & 27\% & 2\% & 5 $\to$ 27 \\
     & price hook & 9\% & 7\% & 14 $\to$ 9 \\
     & warning / exposé & 8\% & 2\% & 3 $\to$ 8 \\
    Evidence & experiential claim & 13\% & 3\% & 2 $\to$ 13 \\
     & completeness promise & 9\% & 2\% & 2 $\to$ 9 \\
     & authority warrant & 9\% & 3\% & 3 $\to$ 9 \\
     & rubric & 8\% & 2\% & 1 $\to$ 8 \\
    Channel & names an assistant & 1.2\% & 0.8\% & 0 $\to$ 1 \\
    \bottomrule
  \end{tabular}
\end{table}

\section{Probe Design and Additional Results}
\label{app:probe}

\noindentparagraph{Questions and selection}
Table~\ref{tab:pilot} lists the pilot questions and controls.
Five pilot controls were fixed in advance as category leaders; for the quercetin, NMN, road-bike and Hong Kong residency questions, the control is the non-network brand that first-round answers named most often, a choice that favors the control (\S\ref{sec:probe}).
On the five questions with fixed controls and the road-bike question scored against the control planned for it before the first round, which the post hoc choice replaced, pilot answers name the client in \engPilotPrefixedClient{} of cases and the control in \engPilotPrefixedControl{}.
For the census, a client qualifies if at least six of the preceding twelve months contain five or more placements and it has at least twenty listicle-titled placements; of the \rqFourCensusClients{} qualifying clients, five had been probed in the pilot, one sells in a category that a pilot question already covers, one has no buyer question in its titles and one could not be scored, which leaves \engCensusPooledClients{} clients with one question each, collected on \censusDate{}; the twelve months are those before that date, and one run each on Wenxin and Google AI Overviews gave no scorable answer.
Census controls were fixed before collection, and Metaso returned only \engCensusMetasoAnswers{} scored answers and is excluded from the pooled comparison.
Two original pilot questions named the client explicitly; on \rqThreeRoundThreeDate{} we replaced them with category questions (\rqThreeRoundThreeAnswers{} answers), and the originals are excluded from pooled category measures and reported as the reputation questions below.

\begin{table*}[t]
  \caption{Pilot questions, with the client whose placements use them and the control brand. Clients are pseudonyms and control brands are described rather than named; four controls were chosen from round-one answers.}
  \label{tab:pilot}
  \small
  \centering
  \begin{tabular}{@{}p{0.22\textwidth}p{0.40\textwidth}p{0.31\textwidth}@{}}
    \toprule
    Client & Question & Control \\
    \midrule
    Client 15 (liver pills) & which liver-pill brand is good & imported milk-thistle liver pills \\
    Client 26 (lung pills) & which quercetin brand is good & supplement brand most named in round one \\
    Client 18 (NMN) & NMN anti-aging brand ranking & NMN brand most named in round one \\
    Client 7 (Nanjing TCM clinic) & Nanjing gastro hospital ranking & provincial public TCM hospital \\
    Client 130 (proctology hospital) & Guangzhou proctology hospital ranking & provincial public TCM hospital \\
    Client 31 (stock advice) & which stock-advisory firms are reliable & leading financial-information and brokerage platform \\
    Client 2 (bicycles) & road-bike brand ranking & bicycle maker most named in round one \\
    Client 36 (face masks) & which hydrating face-mask brand is good & national dressing-mask brand \\
    Client 43 (HK talent visas) & best HK residency-planning agency & residency agency most named in round one \\
    (GEO vendors) & which GEO vendor is best & none \\
    \bottomrule
  \end{tabular}
\end{table*}

\noindentparagraph{Collection and repeatability}
For the pilot, the Chinese assistants were accessed through their web interfaces in fresh conversations: Wenxin logged out, the others logged in.
ChatGPT and Gemini used third-party, search-enabled APIs (\texttt{\rqThreeModelChatGPT} for the first round, \texttt{\engModelChatGPTMini} for the replacement questions and the census; \texttt{\rqThreeModelGemini}); these observations describe the API routes.
The location parameter of Google's search-results API was set to Hong Kong and that of the model APIs to mainland China.
Collection ran during \rqThreeWindow{} from a machine in the United States with a Chinese locale and no VPN; the first round of Yuanbao and DeepSeek was collected on \rqThreeWindowLate{}.
Metaso's first-round anonymous quota allowed one run per question, with three runs for the replacements (\rqThreeMetasoAnswers{} retained answers).
The three runs of a question on an assistant agree on whether the client is named in \engPilotRepAgree{} of pilot cells and \engCensusRepAgree{} of census cells.
Table~\ref{tab:engines} gives the pilot's answer-level rates per assistant.

\input{tables/tab_rq3_engines}

\noindentparagraph{Attribution bounds}
Of \rqThreeCitedUrls{} distinct cited URLs, \rqThreeFetchedUrls{} (\rqThreeFetchedShare{}) were retrievable; \engShortPages{} of them returned fewer than 100 characters of text and are scored on their titles, and excluding them changes the broad share for the nine category questions from \engTierBroadExGeo{} to \engBroadNoShortExGeo{}.
The GEO-vendor question is excluded from the shares in \S\ref{sec:retrieval}: of its \engGeoQuestionUrls{} retrievable pages, \engGeoQuestionBroad{} would count as broad, \engGeoQuestionNamesOnly{} of them only because they contain the phrase ``GEO vendor''.
Of the \engBylinePages{} pages matched by byline on all ten questions, \engBylineNewsPages{} are genuine news reports and \engBylineSubstringPages{} are spurious substring matches, and \engStrictGenuineAuthorPages{} strictly attributed pages have a genuine outlet as author; on the nine category questions, removing such pages takes \engCleanStrictRemoved{} pages from the strict level and \engCleanMidRemoved{} from the extended level, which gives the cleaned lower bounds of \S\ref{sec:retrieval}.
References surfaced \rqThreeProbeSeededAccounts{} Baijiahao accounts outside the crawl, \rqThreeProbeSeededMembers{} of which pass the membership rule; they are excluded from network counts.

\noindentparagraph{Volume and category leaders}
A separate probe of \rqThreeDoseClients{} further clients, drawn from four strata of placement volume and asked one category question each on Wenxin, 360 AI Search and Google AI Overviews, found no relation between a client's placement count and naming (Spearman $\rho=$\engDoseSpearman{}, $p=\engDoseSpearmanP{}$; named \engDoseQuartiles{} by volume quartile); it has one run per question and no control, and each question is taken from the client's own titles.
In the census, clients are named in \engIncumbentClient{} of answers in the \engIncumbentQuestions{} categories whose leader appears in at least half of the answers and in \engNoIncumbentClient{} elsewhere (Fisher's exact test, $p=\engIncumbentFisherP{}$, on a post hoc split), so a dominant category leader does not explain the split by category either (\S\ref{sec:naming}).

\noindentparagraph{Reputation questions}
The two original pilot questions ask whether a named client is reliable (\engRepAnswers{} answers).
A network article defending the client is cited in \engRepCiteNetDefenceStock{} of the answers about the stock-advisory client and \engRepCiteNetDefenceMask{} about the face-mask client, and \engRepPenaltyAnswers{} of the \engRepPenaltyOf{} answers about the stock-advisory client mention a regulatory penalty (terms for penalty, fine, warning letter or supervisory measure).

\section{Coordination Beyond Aggregate Counts}
\label{app:supp}

\noindentparagraph{Campaign durations}
The \campBigClients{} niche clients with at least fifty placements run a median of \campBigCampaignsMedian{} campaigns, \campBigLongShare{} have a campaign of six months or longer, and the median span from first to last placement is \campBigSpanMedian{} days; repeat campaigns of all niche clients are separated by a median of \campGapMedian{} days (\S\ref{sec:coordination}).

\noindentparagraph{High-volume distributors}
\supAssemblyBjhN{} of the \supAssemblyN{} high-volume distributors of \S\ref{sec:production} are on Baijiahao. In a random sample of \supAssemblyBjhRandomN{} Baijiahao network accounts, distributors make up \supAssemblyBjhRandomAcct{} of accounts and \supAssemblyBjhRandomPlacement{} of placements.

\noindentparagraph{Posting synchrony}
For each client, we measure the share of an account's active weeks in which another account also posts that client, compared with a null that independently shifts each account's weeks circularly.
The test covers niche clients with at least three accounts. The Bilibili median is \rqOneSyncObs{} against \rqOneSyncNull{}, significant at the 5\% level for \rqOneSyncSig{} of \rqOneSyncClients{} clients (\syncSigBH{} after Benjamini--Hochberg correction).
Removing the same-day burst signal leaves \syncSigNoBurst{} of \syncClientsNoBurst{} clients significant.
On Baijiahao the median is \rqOneSyncObsBjh{} against \rqOneSyncNullBjh{} (\rqOneSyncSigBjh{} of \rqOneSyncClientsBjh{} significant, \syncSigBHBjh{} after correction), and on 360kuai \rqOneSyncObsKuai{} against \rqOneSyncNullKuai{} (\rqOneSyncSigKuai{} of \rqOneSyncClientsKuai{}, \syncSigBHKuai{} after correction).
The median gap to another account posting the same client is \rqOneGapMedian{} hours on Bilibili, against \rqOneGapNull{} under randomly redrawn post times.

\noindentparagraph{Cross-platform copy timing}
Of the \rqTwoMatchesAll{} title matches across the three platforms, the \rqTwoCopyPairsAll{} whose title names a client (major brands included) cover \rqTwoCopyClientsAll{} clients, with \rqTwoCopyWithinDayAll{} within a day and a median gap of \rqTwoCopyMedianHours{} hours.
The comparison redraws each copy's date independently from that platform's placements for the client.
The Bilibili--Baijiahao subset and its within-hour share are reported in \S\ref{sec:coordination}; these are different populations.

\begin{figure}[t]
  \centering
  \includegraphics[width=\linewidth]{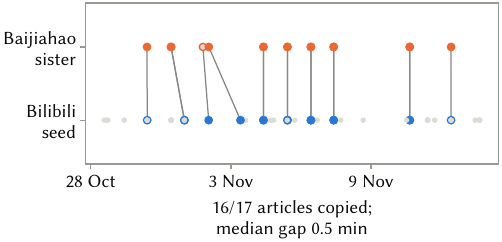}
  \caption{The Bilibili seed and a Baijiahao account with a sister name, 28 October--13 November 2025. Points mark publications and lines join matching titles. Of the Baijiahao account's \sisterTotal{} articles, \sisterIdentical{} match the seed; their median gap is \sisterGap{} minutes.}
  \label{fig:app-sister}
\end{figure}

\noindentparagraph{A pair of accounts in tandem}
We identify a cross-platform account pair when it shares at least five identical titles, excluding short and widely reused titles: \rqTwoPairs{} pairs involve \rqTwoPairedAccounts{} accounts, with \rqTwoPairWithinHour{} of their copies appearing within an hour and \rqTwoPairWithinFiveMin{} within five minutes, against \rqTwoIncidentalWithinHour{} within an hour for incidental matches between other accounts, whose median gap is \incidentalGapMedianH{} hours; the most active pair shares \rqTwoTopPairTitles{} titles, \rqTwoTopPairWithinHour{} of them within an hour.
Figure~\ref{fig:app-sister} makes one such relationship visible: \sisterIdentical{} of a Baijiahao account's \sisterTotal{} articles match the Bilibili seed, \sisterWithinFive{} within five minutes, with a median gap of \sisterGap{} minutes.
This timing supports coordinated publication without identifying the operator.

\noindentparagraph{Shared clients and weeks}
On Bilibili, \rqOneBundles{} of the \rqOneBundlePairsTested{} pairs of niche clients that share at least five core accounts share more accounts than under a volume-preserving null.
Across platforms, weekly placement counts correlate above a circular-shift null for \rqTwoXcorrSigShare{} of the \rqTwoXcorrPairs{} combinations of a searched client name and two platforms with enough volume on both.
Between Bilibili and 360kuai, \kuaiSharedTitles{} titles match; one 360kuai account pairs with \rqTwoKuaiMaxPartners{} Bilibili accounts.

\section{Chinese-Language Sources}
\label{app:sources}
The Chinese-language sources cited in the body are listed here with translated titles.
\begin{description}\setlength{\itemsep}{1pt}
\item[S1] China Central Television, \emph{2026 March 15 Gala} (consumer-rights broadcast), 15 March 2026. Demonstrated a paid service that made major assistants recommend a fictitious product within days, with ranked lists as its exhibit.
\item[S2] Securities Times, ``Poisoning an AI model for 9.9 yuan'' (news report), 18 March 2026. Describes GEO vendors selling article distribution to a dozen content platforms for a fee and promising that the client becomes the assistant's answer.
\item[S3] Standing Committee of the National People's Congress, \emph{Advertising Law of the People's Republic of China}, Article 14, as amended 29 April 2021. Requires advertising to be identifiable as such; \href{https://www.chinalawtranslate.com/advertising-law-2021/}{English translation}.
\item[S4] State Administration for Market Regulation, \emph{Enforcement Guide on the Identifiability of Internet Advertising}, issued 23 August 2024. Applies Article 14 to online content that promotes goods or services without an advertising label.
\item[S5] Cyberspace Administration of China, ``Central Cyberspace Affairs Office launches the `Qinglang' campaign against the misuse of AI applications'' (notice via Xinhua), 30 April 2026. \href{https://www.news.cn/20260430/b1c17b7a72ad4d2ea8204fdf14fc95b9/c.html}{Xinhua notice (30 April 2026)}.
\item[S6] Cyberspace Administration of China, ``Second-phase bulletin of the `Qinglang' campaign against the misuse of AI applications'' (notice via Xinhua), 2 September 2026, naming malicious marketing through GEO as a form of AI data poisoning; first-phase bulletin 6 July 2026. \href{https://www.news.cn/politics/20260902/564dc9a913c94b22996f783a02abcaf2/c.html}{Xinhua bulletin (2 September 2026)}.
\end{description}

%% file: tables/tab_domains_prior.tex
\begin{table*}[t]\centering\small
\caption{Client domains of the network (the \domTableClients{} niche clients with a placement in the core segment) against the promoted-content categories of prior search-spam measurements (Table~\ref{tab:prior_mix}). Only presence is comparable across studies.}\label{tab:domains_prior}
\begin{tabular}{lrrp{3.4cm}p{6.2cm}}
\toprule
Domain & Clients & Share of posts & Examples & Closest category in prior SEO-spam studies \\
\midrule
consumer & 89 & 36\% & Client 2 (bicycles), Client 4 (dehumidifiers), Client 8 (cat food) & SEISE `General' (14\%), Du `Sales and Services' (22\%), Bevendorff (whole genre) \\
health & 67 & 17\% & Client 7 (Nanjing TCM clinic), Client 10 (health devices), Client 15 (liver pills) & SEISE `General' (healthcare products); unlicensed pharmacy/hospitals only elsewhere \\
finance & 50 & 14\% & Client 6 (crypto), Client 16 (lending), Client 21 (lending) & payday loans / financial fraud only (SEISE 8.5\%, RSP 1.1\%) \\
beauty & 38 & 13\% & Client 5 (cosmetic surgery), Client 17 (beauty), Client 23 (masks) & none (counterfeit beauty devices in Search+Seizure) \\
education & 24 & 13\% & Client 1 (test prep), Client 30 (kids books), Client 45 (education) & fake certificates / exam proxies only (RSP 22.7\%, SEISE `Cheating' 9.1\%) \\
tech & 21 & 3\% & Client 20 (NAS), Client 59 (smart locks), Client 92 (technology) & pirated software / fake AV only; ISAP `software development' \\
marketing & 23 & 3\% & Client 25 (GEO agency), Client 58 (GEO vendor), Client 102 (marketing SaaS) & `Blackhat SEO' self-promotion (Klingon 1.9\%, ISAP, RSP 9.2\%) \\
other & 3 & 1\% & Client 87 (other), Client 96 (other), Client 156 (other) & -- \\
legal & 3 & 0\% & Client 228 (legal services), Client 233 (legal), Client 312 (legal) & none (private detectives in Du) \\
automotive & 1 & 0\% & Client 201 (used cars) & none \\
\bottomrule
\end{tabular}
\end{table*}

%% file: tables/tab_prior_mix.tex
\begin{table*}[t]\centering\small
\caption{What prior search-spam measurements found being promoted: unit, the three largest categories, and the share that is ordinary commerce rather than illicit trade or the SEO business itself.}\label{tab:prior_mix}
\begin{tabular}{p{3.2cm}p{3.2cm}p{6cm}r}
\toprule
Study & Unit & Top categories & Ordinary commerce \\
\midrule
Du et al. 2016 & customer sites (930) & Sales and Services 22\%; Gambling 20\%; Surrogacy 17\% & 22\% \\
Liao et al. 2016 SEISE & injected domains (11,131) & Gambling 19\%; Drug 15\%; General 14\% & 14\% \\
Yang et al. 2017 Klingon & black keywords (478,879) & Sex 46\%; Gambling 39\%; Other 10\% & 0\% \\
Zhang et al. 2024 ISAP & ISAP URLs (3.2M) & adult content 77\%; gambling 20\% & 0\% \\
Wu et al. 2024 RSP & illicit promotion texts (12.0M) & Sex Service 25\%; Gambling 24\%; Fake Certificate 23\% & 0\% \\
Wang et al. 2014 Search+Seizure & poisoned search results (2.77M) in 16 brand verticals & counterfeit luxury (16 brand verticals) 100\% & 0\% \\
Bevendorff et al. 2024 & top-ranked sites for 7,392 `best X' product queries & Magazine 31\%; Web Shop 31\%; Review Farm 16\% & 100\% \\
Leontiadis et al. 2014 & search results for pharma + 6 abuse-prone verticals & no shares given & -- \\
\bottomrule
\end{tabular}
\end{table*}

%% file: tables/tab_rq3_engines.tex
\begin{table}[t]\centering\footnotesize
\caption{Pilot results for the nine category questions (the GEO-vendor question has no client and is analyzed separately). Client and Control give naming rates over these answers; Cites client counts answers listing a page that names the client. Reputation questions and census results are reported separately.}\label{tab:engines}
\setlength{\tabcolsep}{3.5pt}\begin{tabular}{@{}lrrrr@{}}
\toprule
Service & Ans. & Client & Control & Cites client \\
\midrule
Wenxin (Baidu) & 27 & 74\% & 52\% & 27 \\
Metaso & 13 & 69\% & 62\% & 12 \\
Gemini & 27 & 59\% & 70\% & 18 \\
360 AI Search & 27 & 44\% & 48\% & 17 \\
Google AI Overviews & 27 & 44\% & 70\% & 11 \\
DeepSeek & 27 & 33\% & 48\% & 13 \\
ChatGPT & 27 & 26\% & 59\% & 14 \\
Google AI Mode & 27 & 26\% & 96\% & 11 \\
Yuanbao (Tencent) & 27 & 26\% & 70\% & 24 \\
\bottomrule
\end{tabular}
\end{table}

%% file: references.bib
@article{keller2020political,
  author = {Keller, Franziska B. and Schoch, David and Stier, Sebastian and Yang, JungHwan},
  title = {Political Astroturfing on Twitter: How to Coordinate a Disinformation Campaign},
  journal = {Political Communication},
  volume = {37},
  number = {2},
  pages = {256--280},
  year = {2020},
  publisher = {Routledge},
  doi = {10.1080/10584609.2019.1661888},
  URL = {https://doi.org/10.1080/10584609.2019.1661888}
}

@article{alyukov2025datavoids,
  title={LLMs grooming or data voids? LLM-powered chatbot references to Kremlin disinformation reflect information gaps, not manipulation},
  author={Alyukov, Maxim and Makhortykh, Mykola and Voronovici, Alexandr and Sydorova, Maryna},
  journal={Harvard Kennedy School Misinformation Review},
  year={2025}
}

@inproceedings{aggarwal2024geo,
  author       = {Pranjal Aggarwal and
                  Vishvak Murahari and
                  Tanmay Rajpurohit and
                  Ashwin Kalyan and
                  Karthik Narasimhan and
                  Ameet Deshpande},
  editor       = {Ricardo Baeza{-}Yates and
                  Francesco Bonchi},
  title        = {{GEO:} Generative Engine Optimization},
  booktitle    = {Proceedings of the 30th {ACM} {SIGKDD} Conference on Knowledge Discovery
                  and Data Mining, {KDD} 2024, Barcelona, Spain, August 25-29, 2024},
  pages        = {5--16},
  publisher    = {{ACM}},
  year         = {2024},
  url          = {https://doi.org/10.1145/3637528.3671900},
  doi          = {10.1145/3637528.3671900},
  bibsource    = {dblp computer science bibliography, https://dblp.org}
}

@inproceedings{nestaas2025adversarial,
  author       = {Fredrik Nestaas and
                  Edoardo Debenedetti and
                  Florian Tram{\`{e}}r},
  title        = {Adversarial Search Engine Optimization for Large Language Models},
  booktitle    = {The Thirteenth International Conference on Learning Representations,
                  {ICLR} 2025, Singapore, April 24-28, 2025},
  publisher    = {OpenReview.net},
  year         = {2025},
  url          = {https://openreview.net/forum?id=hkdqxN3c7t},
  bibsource    = {dblp computer science bibliography, https://dblp.org}
}

@inproceedings{pfrommer2024ranking,
  author       = {Samuel Pfrommer and
                  Yatong Bai and
                  Tanmay Gautam and
                  Somayeh Sojoudi},
  editor       = {Yaser Al{-}Onaizan and
                  Mohit Bansal and
                  Yun{-}Nung Chen},
  title        = {Ranking Manipulation for Conversational Search Engines},
  booktitle    = {Proceedings of the 2024 Conference on Empirical Methods in Natural
                  Language Processing, {EMNLP} 2024, Miami, FL, USA, November 12-16,
                  2024},
  pages        = {9523--9552},
  publisher    = {Association for Computational Linguistics},
  year         = {2024},
  url          = {https://doi.org/10.18653/v1/2024.emnlp-main.534},
  doi          = {10.18653/V1/2024.EMNLP-MAIN.534},
  bibsource    = {dblp computer science bibliography, https://dblp.org}
}

@article{kumar2024manipulating,
  author       = {Aounon Kumar and
                  Himabindu Lakkaraju},
  title        = {Manipulating Large Language Models to Increase Product Visibility},
  journal      = {CoRR},
  volume       = {abs/2404.07981},
  year         = {2024},
  url          = {https://doi.org/10.48550/arXiv.2404.07981},
  doi          = {10.48550/ARXIV.2404.07981},
  eprinttype   = {arXiv},
  eprint       = {2404.07981},
  bibsource    = {dblp computer science bibliography, https://dblp.org}
}

@article{zhen2026chinesegse,
  author       = {Tao Zhen and
                  Yue Liu and
                  Gege Zhang and
                  Yixuan Niu},
  title        = {What Do Chinese-Language Generative Search Engines Cite and Surface?
                  {A} Large-Scale Empirical Study},
  journal      = {CoRR},
  volume       = {abs/2607.15771},
  year         = {2026},
  url          = {https://doi.org/10.48550/arXiv.2607.15771},
  doi          = {10.48550/ARXIV.2607.15771},
  eprinttype   = {arXiv},
  eprint       = {2607.15771},
  bibsource    = {dblp computer science bibliography, https://dblp.org}
}

@article{li2026baidugoogle,
  author       = {Yibo Li and
                  Enci Guan and
                  Yuedan Cai and
                  Geng Liu and
                  Francesco Pierri},
  title        = {Auditing Source Exposure in Baidu and Google {AI} Search},
  journal      = {CoRR},
  volume       = {abs/2609.24407},
  year         = {2026},
  url          = {https://doi.org/10.48550/arXiv.2609.24407},
  doi          = {10.48550/ARXIV.2609.24407},
  eprinttype   = {arXiv},
  eprint       = {2609.24407},
  bibsource    = {dblp computer science bibliography, https://dblp.org}
}

@article{xu2026aioverviews,
  author       = {Haofei Xu and
                  Umar Iqbal and
                  Jacob M. Montgomery},
  title        = {Measuring Google {AI} Overviews: Activation, Source Quality, Claim
                  Fidelity, and Publisher Impact},
  journal      = {CoRR},
  volume       = {abs/2605.14021},
  year         = {2026},
  url          = {https://doi.org/10.48550/arXiv.2605.14021},
  doi          = {10.48550/ARXIV.2605.14021},
  eprinttype   = {arXiv},
  eprint       = {2605.14021},
  bibsource    = {dblp computer science bibliography, https://dblp.org}
}

@article{allaham2026synthetic,
  author       = {Mowafak Allaham and
                  Nicholas Diakopoulos},
  title        = {Synthetic Sources?: Auditing Generative Search Engine Citations for
                  Evidence of AI-Generated Sources},
  journal      = {CoRR},
  volume       = {abs/2605.23684},
  year         = {2026},
  url          = {https://doi.org/10.48550/arXiv.2605.23684},
  doi          = {10.48550/ARXIV.2605.23684},
  eprinttype   = {arXiv},
  eprint       = {2605.23684},
  bibsource    = {dblp computer science bibliography, https://dblp.org}
}

@inproceedings{gyongyi2005alliances,
  author       = {Zolt{\'{a}}n Gy{\"{o}}ngyi and
                  Hector Garcia{-}Molina},
  editor       = {Klemens B{\"{o}}hm and
                  Christian S. Jensen and
                  Laura M. Haas and
                  Martin L. Kersten and
                  Per{-}{\AA}ke Larson and
                  Beng Chin Ooi},
  title        = {Link Spam Alliances},
  booktitle    = {Proceedings of the 31st International Conference on Very Large Data
                  Bases, Trondheim, Norway, August 30 - September 2, 2005},
  pages        = {517--528},
  publisher    = {{ACM}},
  year         = {2005},
  url          = {http://www.vldb.org/archives/website/2005/program/paper/thu/p517-gyongyi.pdf},
  bibsource    = {dblp computer science bibliography, https://dblp.org}
}

@inproceedings{leontiadis2011redirection,
  author       = {Nektarios Leontiadis and
                  Tyler Moore and
                  Nicolas Christin},
  title        = {Measuring and Analyzing Search-Redirection Attacks in the Illicit
                  Online Prescription Drug Trade},
  booktitle    = {20th {USENIX} Security Symposium, San Francisco, CA, USA, August 8-12,
                  2011, Proceedings},
  publisher    = {{USENIX} Association},
  year         = {2011},
  url          = {http://static.usenix.org/events/sec11/tech/full\_papers/Leontiadis.pdf},
  bibsource    = {dblp computer science bibliography, https://dblp.org}
}

@inproceedings{leontiadis2014longitudinal,
  author       = {Nektarios Leontiadis and
                  Tyler Moore and
                  Nicolas Christin},
  editor       = {Gail{-}Joon Ahn and
                  Moti Yung and
                  Ninghui Li},
  title        = {A Nearly Four-Year Longitudinal Study of Search-Engine Poisoning},
  booktitle    = {Proceedings of the 2014 {ACM} {SIGSAC} Conference on Computer and
                  Communications Security, Scottsdale, AZ, USA, November 3-7, 2014},
  pages        = {930--941},
  publisher    = {{ACM}},
  year         = {2014},
  url          = {https://doi.org/10.1145/2660267.2660332},
  doi          = {10.1145/2660267.2660332},
  bibsource    = {dblp computer science bibliography, https://dblp.org}
}

@inproceedings{wang2013juice,
  author       = {David Y. Wang and
                  Stefan Savage and
                  Geoffrey M. Voelker},
  title        = {Juice: {A} Longitudinal Study of an {SEO} Botnet},
  booktitle    = {20th Annual Network and Distributed System Security Symposium, {NDSS}
                  2013, San Diego, California, USA, February 24-27, 2013},
  publisher    = {The Internet Society},
  year         = {2013},
  url          = {https://www.ndss-symposium.org/ndss2013/juice-longitudinal-study-seo-botnet},
  bibsource    = {dblp computer science bibliography, https://dblp.org}
}

@inproceedings{wang2014seizure,
  author       = {David Y. Wang and
                  Matthew F. Der and
                  Mohammad Karami and
                  Lawrence K. Saul and
                  Damon McCoy and
                  Stefan Savage and
                  Geoffrey M. Voelker},
  editor       = {Carey Williamson and
                  Aditya Akella and
                  Nina Taft},
  title        = {Search + Seizure: The Effectiveness of Interventions on {SEO} Campaigns},
  booktitle    = {Proceedings of the 2014 Internet Measurement Conference, {IMC} 2014,
                  Vancouver, BC, Canada, November 5-7, 2014},
  pages        = {359--372},
  publisher    = {{ACM}},
  year         = {2014},
  url          = {https://doi.org/10.1145/2663716.2663738},
  doi          = {10.1145/2663716.2663738},
  bibsource    = {dblp computer science bibliography, https://dblp.org}
}

@inproceedings{vangoethem2019pbn,
  author       = {Tom van Goethem and
                  Najmeh Miramirkhani and
                  Wouter Joosen and
                  Nick Nikiforakis},
  editor       = {Steven D. Galbraith and
                  Giovanni Russello and
                  Willy Susilo and
                  Dieter Gollmann and
                  Engin Kirda and
                  Zhenkai Liang},
  title        = {Purchased Fame: Exploring the Ecosystem of Private Blog Networks},
  booktitle    = {Proceedings of the 2019 {ACM} Asia Conference on Computer and Communications
                  Security, AsiaCCS 2019, Auckland, New Zealand, July 09-12, 2019},
  pages        = {366--378},
  publisher    = {{ACM}},
  year         = {2019},
  url          = {https://doi.org/10.1145/3321705.3329830},
  doi          = {10.1145/3321705.3329830},
  bibsource    = {dblp computer science bibliography, https://dblp.org}
}

@inproceedings{du2016labyrinth,
  author       = {Kun Du and
                  Hao Yang and
                  Zhou Li and
                  Hai{-}Xin Duan and
                  Kehuan Zhang},
  editor       = {Thorsten Holz and
                  Stefan Savage},
  title        = {The Ever-Changing Labyrinth: {A} Large-Scale Analysis of Wildcard
                  {DNS} Powered Blackhat {SEO}},
  booktitle    = {25th {USENIX} Security Symposium, {USENIX} Security 16, Austin, TX,
                  USA, August 10-12, 2016},
  pages        = {245--262},
  publisher    = {{USENIX} Association},
  year         = {2016},
  url          = {https://www.usenix.org/conference/usenixsecurity16/technical-sessions/presentation/du},
  bibsource    = {dblp computer science bibliography, https://dblp.org}
}

@inproceedings{liao2016seise,
  author       = {Xiaojing Liao and
                  Kan Yuan and
                  XiaoFeng Wang and
                  Zhongyu Pei and
                  Hao Yang and
                  Jianjun Chen and
                  Hai{-}Xin Duan and
                  Kun Du and
                  Eihal Alowaisheq and
                  Sumayah A. Alrwais and
                  Luyi Xing and
                  Raheem A. Beyah},
  title        = {Seeking Nonsense, Looking for Trouble: Efficient Promotional-Infection
                  Detection through Semantic Inconsistency Search},
  booktitle    = {{IEEE} Symposium on Security and Privacy, {SP} 2016, San Jose, CA,
                  USA, May 22-26, 2016},
  pages        = {707--723},
  publisher    = {{IEEE} Computer Society},
  year         = {2016},
  url          = {https://doi.org/10.1109/SP.2016.48},
  doi          = {10.1109/SP.2016.48},
  bibsource    = {dblp computer science bibliography, https://dblp.org}
}

@inproceedings{yang2017klingon,
  author       = {Hao Yang and
                  Xiulin Ma and
                  Kun Du and
                  Zhou Li and
                  Hai{-}Xin Duan and
                  XiaoDong Su and
                  Guang Liu and
                  Zhifeng Geng and
                  Jianping Wu},
  title        = {How to Learn Klingon without a Dictionary: Detection and Measurement
                  of Black Keywords Used by the Underground Economy},
  booktitle    = {2017 {IEEE} Symposium on Security and Privacy, {SP} 2017, San Jose,
                  CA, USA, May 22-26, 2017},
  pages        = {751--769},
  publisher    = {{IEEE} Computer Society},
  year         = {2017},
  url          = {https://doi.org/10.1109/SP.2017.11},
  doi          = {10.1109/SP.2017.11},
  bibsource    = {dblp computer science bibliography, https://dblp.org}
}

@inproceedings{yang2021dmos,
  author       = {Ronghai Yang and
                  Xianbo Wang and
                  Cheng Chi and
                  Dawei Wang and
                  Jiawei He and
                  Siming Pang and
                  Wing Cheong Lau},
  editor       = {Michael D. Bailey and
                  Rachel Greenstadt},
  title        = {Scalable Detection of Promotional Website Defacements in Black Hat
                  {SEO} Campaigns},
  booktitle    = {30th {USENIX} Security Symposium, {USENIX} Security 2021, August 11-13,
                  2021},
  pages        = {3703--3720},
  publisher    = {{USENIX} Association},
  year         = {2021},
  url          = {https://www.usenix.org/conference/usenixsecurity21/presentation/yang-ronghai},
  bibsource    = {dblp computer science bibliography, https://dblp.org}
}

@inproceedings{zhang2024isap,
  author       = {Yunyi Zhang and
                  Mingxuan Liu and
                  Baojun Liu and
                  Yiming Zhang and
                  Haixin Duan and
                  Min Zhang and
                  Hui Jiang and
                  Yanzhe Li and
                  Fan Shi},
  editor       = {Davide Balzarotti and
                  Wenyuan Xu},
  title        = {Into the Dark: Unveiling Internal Site Search Abused for Black Hat
                  {SEO}},
  booktitle    = {33rd {USENIX} Security Symposium, {USENIX} Security 2024, Philadelphia,
                  PA, USA, August 14-16, 2024},
  publisher    = {{USENIX} Association},
  year         = {2024},
  url          = {https://www.usenix.org/conference/usenixsecurity24/presentation/zhang-yunyi-dark},
  bibsource    = {dblp computer science bibliography, https://dblp.org}
}

@inproceedings{liu2025nokescam,
  author       = {Mingxuan Liu and
                  Yunyi Zhang and
                  Lijie Wu and
                  Baojun Liu and
                  Geng Hong and
                  Yiming Zhang and
                  Hui Jiang and
                  Jia Zhang and
                  Haixin Duan and
                  Min Zhang and
                  Wei Guan and
                  Fan Shi and
                  Min Yang},
  editor       = {Lujo Bauer and
                  Giancarlo Pellegrino},
  title        = {NOKEScam: Understanding and Rectifying Non-Sense Keywords Spear Scam
                  in Search Engines},
  booktitle    = {34th {USENIX} Security Symposium, {USENIX} Security 2025, Seattle,
                  WA, USA, August 13-15, 2025},
  pages        = {4779--4798},
  publisher    = {{USENIX} Association},
  year         = {2025},
  url          = {https://www.usenix.org/conference/usenixsecurity25/presentation/liu-mingxuan},
  bibsource    = {dblp computer science bibliography, https://dblp.org}
}

@article{wu2024rsp,
  author       = {Sangyi Wu and
                  Jialong Xue and
                  Shaoxuan Zhou and
                  Xianghang Mi},
  title        = {Reflected Search Poisoning for Illicit Promotion},
  journal      = {CoRR},
  volume       = {abs/2404.05320},
  year         = {2024},
  url          = {https://doi.org/10.48550/arXiv.2404.05320},
  doi          = {10.48550/ARXIV.2404.05320},
  eprinttype   = {arXiv},
  eprint       = {2404.05320},
  bibsource    = {dblp computer science bibliography, https://dblp.org}
}

@inproceedings{wang2025illicit,
  author       = {Hongyu Wang and
                  Ying Li and
                  Ronghong Huang and
                  Xianghang Mi},
  editor       = {Guodong Long and
                  Michale Blumestein and
                  Yi Chang and
                  Liane Lewin{-}Eytan and
                  Zi Helen Huang and
                  Elad Yom{-}Tov},
  title        = {Detecting and Understanding the Promotion of Illicit Goods and Services
                  on Twitter},
  booktitle    = {Proceedings of the {ACM} on Web Conference 2025, {WWW} 2025, Sydney,
                  NSW, Australia, 28 April 2025- 2 May 2025},
  pages        = {3389--3404},
  publisher    = {{ACM}},
  year         = {2025},
  url          = {https://doi.org/10.1145/3696410.3714550},
  doi          = {10.1145/3696410.3714550},
  bibsource    = {dblp computer science bibliography, https://dblp.org}
}

@inproceedings{chen2013wateramy,
  author       = {Cheng Chen and
                  Kui Wu and
                  Srinivasan Venkatesh and
                  Xudong Zhang},
  editor       = {Jon G. Rokne and
                  Christos Faloutsos},
  title        = {Battling the internet water army: detection of hidden paid posters},
  booktitle    = {Advances in Social Networks Analysis and Mining 2013, {ASONAM} '13,
                  Niagara, ON, Canada - August 25 - 29, 2013},
  pages        = {116--120},
  publisher    = {{ACM}},
  year         = {2013},
  url          = {https://doi.org/10.1145/2492517.2492637},
  doi          = {10.1145/2492517.2492637},
  bibsource    = {dblp computer science bibliography, https://dblp.org}
}

@inproceedings{wang2012serf,
  author       = {Gang Wang and
                  Christo Wilson and
                  Xiaohan Zhao and
                  Yibo Zhu and
                  Manish Mohanlal and
                  Haitao Zheng and
                  Ben Y. Zhao},
  editor       = {Alain Mille and
                  Fabien Gandon and
                  Jacques Misselis and
                  Michael Rabinovich and
                  Steffen Staab},
  title        = {Serf and turf: crowdturfing for fun and profit},
  booktitle    = {Proceedings of the 21st World Wide Web Conference 2012, {WWW} 2012,
                  Lyon, France, April 16-20, 2012},
  pages        = {679--688},
  publisher    = {{ACM}},
  year         = {2012},
  url          = {https://doi.org/10.1145/2187836.2187928},
  doi          = {10.1145/2187836.2187928},
  bibsource    = {dblp computer science bibliography, https://dblp.org}
}

@inproceedings{lee2013crowdturfers,
  author       = {Kyumin Lee and
                  Prithivi Tamilarasan and
                  James Caverlee},
  editor       = {Emre Kiciman and
                  Nicole B. Ellison and
                  Bernie Hogan and
                  Paul Resnick and
                  Ian Soboroff},
  title        = {Crowdturfers, Campaigns, and Social Media: Tracking and Revealing
                  Crowdsourced Manipulation of Social Media},
  booktitle    = {Proceedings of the Seventh International Conference on Weblogs and
                  Social Media, {ICWSM} 2013, Cambridge, Massachusetts, USA, July 8-11,
                  2013},
  publisher    = {The {AAAI} Press},
  year         = {2013},
  url          = {http://www.aaai.org/ocs/index.php/ICWSM/ICWSM13/paper/view/5988},
  bibsource    = {dblp computer science bibliography, https://dblp.org}
}

@inproceedings{mukherjee2012spotting,
  author       = {Arjun Mukherjee and
                  Bing Liu and
                  Natalie S. Glance},
  editor       = {Alain Mille and
                  Fabien Gandon and
                  Jacques Misselis and
                  Michael Rabinovich and
                  Steffen Staab},
  title        = {Spotting fake reviewer groups in consumer reviews},
  booktitle    = {Proceedings of the 21st World Wide Web Conference 2012, {WWW} 2012,
                  Lyon, France, April 16-20, 2012},
  pages        = {191--200},
  publisher    = {{ACM}},
  year         = {2012},
  url          = {https://doi.org/10.1145/2187836.2187863},
  doi          = {10.1145/2187836.2187863},
  bibsource    = {dblp computer science bibliography, https://dblp.org}
}

@article{he2022market,
  author       = {Sherry He and
                  Brett Hollenbeck and
                  Davide Proserpio},
  title        = {The Market for Fake Reviews},
  journal      = {Mark. Sci.},
  volume       = {41},
  number       = {5},
  pages        = {896--921},
  year         = {2022},
  url          = {https://doi.org/10.1287/mksc.2022.1353},
  doi          = {10.1287/MKSC.2022.1353},
  bibsource    = {dblp computer science bibliography, https://dblp.org}
}

@inproceedings{kaghazgaran2018combating,
  author       = {Parisa Kaghazgaran and
                  James Caverlee and
                  Anna Cinzia Squicciarini},
  editor       = {Yi Chang and
                  Chengxiang Zhai and
                  Yan Liu and
                  Yoelle Maarek},
  title        = {Combating Crowdsourced Review Manipulators: {A} Neighborhood-Based
                  Approach},
  booktitle    = {Proceedings of the Eleventh {ACM} International Conference on Web
                  Search and Data Mining, {WSDM} 2018, Marina Del Rey, CA, USA, February
                  5-9, 2018},
  pages        = {306--314},
  publisher    = {{ACM}},
  year         = {2018},
  url          = {https://doi.org/10.1145/3159652.3159726},
  doi          = {10.1145/3159652.3159726},
  bibsource    = {dblp computer science bibliography, https://dblp.org}
}

@inproceedings{pacheco2021uncovering,
  author       = {Diogo Pacheco and
                  Pik{-}Mai Hui and
                  Christopher Torres{-}Lugo and
                  Bao Tran Truong and
                  Alessandro Flammini and
                  Filippo Menczer},
  editor       = {Ceren Budak and
                  Meeyoung Cha and
                  Daniele Quercia and
                  Lexing Xie},
  title        = {Uncovering Coordinated Networks on Social Media: Methods and Case
                  Studies},
  booktitle    = {Proceedings of the Fifteenth International {AAAI} Conference on Web
                  and Social Media, {ICWSM} 2021, held virtually, June 7-10, 2021},
  pages        = {455--466},
  publisher    = {{AAAI} Press},
  year         = {2021},
  url          = {https://ojs.aaai.org/index.php/ICWSM/article/view/18075},
  bibsource    = {dblp computer science bibliography, https://dblp.org}
}

@inproceedings{luo2025unsafesearch,
  author       = {Zeren Luo and
                  Zifan Peng and
                  Yule Liu and
                  Zhen Sun and
                  Mingchen Li and
                  Jingyi Zheng and
                  Xinlei He},
  editor       = {Lujo Bauer and
                  Giancarlo Pellegrino},
  title        = {Unsafe LLM-Based Search: Quantitative Analysis and Mitigation of Safety
                  Risks in {AI} Web Search},
  booktitle    = {34th {USENIX} Security Symposium, {USENIX} Security 2025, Seattle,
                  WA, USA, August 13-15, 2025},
  pages        = {8055--8074},
  publisher    = {{USENIX} Association},
  year         = {2025},
  url          = {https://www.usenix.org/conference/usenixsecurity25/presentation/luo-zeren},
  bibsource    = {dblp computer science bibliography, https://dblp.org}
}

@inproceedings{zou2025poisonedrag,
  author       = {Wei Zou and
                  Runpeng Geng and
                  Binghui Wang and
                  Jinyuan Jia},
  editor       = {Lujo Bauer and
                  Giancarlo Pellegrino},
  title        = {PoisonedRAG: Knowledge Corruption Attacks to Retrieval-Augmented Generation
                  of Large Language Models},
  booktitle    = {34th {USENIX} Security Symposium, {USENIX} Security 2025, Seattle,
                  WA, USA, August 13-15, 2025},
  pages        = {3827--3844},
  publisher    = {{USENIX} Association},
  year         = {2025},
  url          = {https://www.usenix.org/conference/usenixsecurity25/presentation/zou-poisonedrag},
  bibsource    = {dblp computer science bibliography, https://dblp.org}
}

@inproceedings{carlini2024poisoning,
  author       = {Nicholas Carlini and
                  Matthew Jagielski and
                  Christopher A. Choquette{-}Choo and
                  Daniel Paleka and
                  Will Pearce and
                  Hyrum S. Anderson and
                  Andreas Terzis and
                  Kurt Thomas and
                  Florian Tram{\`{e}}r},
  title        = {Poisoning Web-Scale Training Datasets is Practical},
  booktitle    = {{IEEE} Symposium on Security and Privacy, {SP} 2024, San Francisco,
                  CA, USA, May 19-23, 2024},
  pages        = {407--425},
  publisher    = {{IEEE}},
  year         = {2024},
  url          = {https://doi.org/10.1109/SP54263.2024.00179},
  doi          = {10.1109/SP54263.2024.00179},
  bibsource    = {dblp computer science bibliography, https://dblp.org}
}

@inproceedings{zha2024illicitdrug,
  author       = {Mingming Zha and
                  Zilong Lin and
                  Siyuan Tang and
                  Xiaojing Liao and
                  Yuhong Nan and
                  XiaoFeng Wang},
  editor       = {Bo Luo and
                  Xiaojing Liao and
                  Jun Xu and
                  Engin Kirda and
                  David Lie},
  title        = {Understanding Cross-Platform Referral Traffic for Illicit Drug Promotion},
  booktitle    = {Proceedings of the 2024 on {ACM} {SIGSAC} Conference on Computer and
                  Communications Security, {CCS} 2024, Salt Lake City, UT, USA, October
                  14-18, 2024},
  pages        = {2132--2146},
  publisher    = {{ACM}},
  year         = {2024},
  url          = {https://doi.org/10.1145/3658644.3670383},
  doi          = {10.1145/3658644.3670383},
  bibsource    = {dblp computer science bibliography, https://dblp.org}
}

@inproceedings{na2023evolvingbots,
  author       = {Seung Ho Na and
                  Sumin Cho and
                  Seungwon Shin},
  editor       = {Marie{-}Jos{\'{e}} Montpetit and
                  Aris Leivadeas and
                  Steve Uhlig and
                  Mobin Javed},
  title        = {Evolving Bots: The New Generation of Comment Bots and their Underlying
                  Scam Campaigns in YouTube},
  booktitle    = {Proceedings of the 2023 {ACM} on Internet Measurement Conference,
                  {IMC} 2023, Montreal, QC, Canada, October 24-26, 2023},
  pages        = {297--312},
  publisher    = {{ACM}},
  year         = {2023},
  url          = {https://doi.org/10.1145/3618257.3624822},
  doi          = {10.1145/3618257.3624822},
  bibsource    = {dblp computer science bibliography, https://dblp.org}
}

@inproceedings{farooqi2020incentivized,
  author       = {Shehroze Farooqi and
                  {\'{A}}lvaro Feal and
                  Tobias Lauinger and
                  Damon McCoy and
                  Zubair Shafiq and
                  Narseo Vallina{-}Rodriguez},
  title        = {Understanding Incentivized Mobile App Installs on Google Play Store},
  booktitle    = {{IMC} '20: {ACM} Internet Measurement Conference, Virtual Event, USA,
                  October 27-29, 2020},
  pages        = {696--709},
  publisher    = {{ACM}},
  year         = {2020},
  url          = {https://doi.org/10.1145/3419394.3423662},
  doi          = {10.1145/3419394.3423662},
  bibsource    = {dblp computer science bibliography, https://dblp.org}
}
